\documentclass[structabstract]{aa}   
\usepackage{graphicx}
\usepackage{txfonts}
\usepackage{natbib}

\newcommand{\HI}{\mbox{H\,{\sc i}}}
\newcommand{\HII}{\mbox{H\,{\sc ii}}}

\begin{document}
   \title{The GALEX Ultraviolet Virgo Cluster Survey (GUViCS). II. Constraints on star formation in ram-pressure stripped gas}


   \author{S. Boissier\inst{1}, A. Boselli\inst{1}, P.-A. Duc\inst{2}, L. Cortese\inst{3}, W. van Driel\inst{4}, S. Heinis\inst{1}, E. Voyer\inst{1}, O. Cucciati\inst{5}, L. Ferrarese\inst{6}, P. C\^ot\'e\inst{6}, J.-C. Cuillandre\inst{7}, S. D. J. Gwyn\inst{6}, \and S. Mei\inst{4,8,9}
 	}
   \institute{Aix Marseille Universit\'e, CNRS, LAM (Laboratoire d'Astrophysique de Marseille) UMR 7326, 13388, Marseille, France
         \and
             Laboratoire AIM Paris-Saclay, CEA/IRFU/SAp, CNRS/INSU, Universit\'e Paris Diderot, 91191 Gif-sur-Yvette Cedex, France
         \and
             European Southern Observatory, Karl Schwarzschild Str. 2, 85748 Garching bei M\"unchen, Germany
	       \and
             GEPI, Observatoire de Paris, CNRS, Universit\'e Paris Diderot, 5 place Jules Janssen, 92190 Meudon, France
         \and
	           INAF-Osservatorio Astronomico di Trieste - Via Tiepolo 11, I-34143 Trieste, Italy
       	\and
	          Herzberg Institute of Astrophysics, National Research Council of Canada, Victoria, BC, V9E 2E7, Canada 
	      \and
	          Canada-France-Hawaii Telescope Corporation, Kamuela, HI 96743, USA
        \and
            Universit\'e Paris Denis Diderot, 75205 Paris Cedex 13, France
        \and
            IPAC - California Institute of Technology, Pasadena, CA 91125, USA
             }

   \date{submitted - 2012}

 
  \abstract
   {Several galaxies in the Virgo cluster are known to have large \HI\ gas tails related to a recent ram-pressure stripping event. The Virgo cluster has been extensively observed at 1539 \AA\ in the far-ultraviolet for the GALEX Ultraviolet Virgo Cluster Survey (GUViCS), and in the optical for the Next Generation Virgo Survey (NGVS), allowing a study of the stellar emission potentially associated with the gas tails of 8 cluster members. On the theoretical side, models of ram-pressure stripping events have started to include the physics of star formation.}
{We aim to provide quantitative constraints on the amount of star formation taking place in the ram-pressure stripped gas, mainly on the basis of the far-UV emission found in the GUViCS images in relation with the gas content of the tails.}
{We have performed three comparisons of the young stars' emission with the gas column density: visual, pixel-by-pixel and global. We have compared our results to other observational and theoretical studies.}
{We find that the level of star formation taking place in the gas stripped from galaxies by ram-pressure is low with respect to the available amount of gas. Star formation is lower by at least a factor 10 compared to the predictions of the Schmidt Law as determined in regular spiral galaxy disks. It is also lower than measured in dwarfs galaxies and the outer regions of spirals, and than predicted by some numerical simulations. We provide constraints on the star formation efficiency in the ram-pressure stripped gas tails, and compare these with current models.}
  {}

   \keywords{Galaxies: clusters: general -- Galaxies: clusters: Virgo -- Galaxies: star formation -- Galaxies: interactions}

\authorrunning{Boissier et al.}
\titlerunning{star formation in ram-pressure stripped galaxies}

   \maketitle
%

\section{Introduction}

The environment can profoundly alter the evolution and nature of galaxies \citep[see e.g. the review by][]{bosellirev}.
In clusters, ram-pressure stripping \citep{gunn72} is one of the major processes whereby gas is removed from the body of galaxies, by the dynamical pressure due to their fast motion through the hot intra-cluster medium (ICM). Its effect on the gas thus depends mainly on the density of the ICM and the relative velocity between the ICM and the galaxy.
In the Virgo cluster in particular, the observation of long, one-sided \HI\ tails without stellar counterparts has led several studies to suggest that many of its galaxies are presently undergoing ram-pressure stripping \citep[e.g.][]{vollmer03,vollmer04,oosterloo05,chung07,chung09}.
Other galaxies show signs of a past ram-pressure stripping event, such as a normal stellar disk which is truncated in \HI, with relatively smaller \HI\ disks being found closer to the Virgo cluster core \citep{cayatte90}.
Many models confirmed that ram-pressure stripping can cause profound transformations in galaxies \citep[e.g.][]{quilis00,schulz01,vollmer01}, such as star-forming dwarfs becoming gas-poor dwarf ellipticals \citep[e.g.][]{boselli08,bobo08}, or making spirals gas deficient \citep[e.g.][]{abadi99,roediger07}, with stellar populations in agreement with observations \citep{boselli06}.

In the case of ram-pressure stripping, the fate of the galaxy may be easier to understand than the fate of the stripped gas itself, due to the complexity of its physics (turbulence, gravity, dynamic pressure, viscosity, cooling, mixing, star formation, feedback...) and the possible dependence on many parameters (galaxy mass and structure, ICM density and temperature, relative velocity, inclination...).
Although attempts were made to model the physics of the stripped gas \citep{roediger05,kapferer09,tonnesen12} it is not certain that all relevant processes have been correctly accounted for. The fate of the gas is not yet clear: part of it may fall back to the galaxy \citep{vollmer01,roediger05}, but various models predict different levels of star formation in the stripped tails \citep{kapferer09,tonnesen12,vollmer12ngc4330}.
Observationally, there are indeed indications of rare (or low-level) star formation in stripped material in rich clusters \citep{cortese07,smith10,sun10,yoshida12} and in a few Virgo cluster galaxies \citep{kenney99,abramson11,fumagalli11}.
Although in \object{VCC 1249} stellar population synthesis models suggest in-situ star formation from gas removed from the galaxy \citep{arrigoni12} the galaxy is undergoing both ram-pressure stripping and tidal interactions. 
Few of these studies include actual constraints on the \HI\ gas distribution itself, except those for Virgo cluster galaxies such as \citet{kenney99,abramson11} or the recent work of \citet{vollmer12sfe} who applied methods similar to ours to 12 Virgo cluster galaxies (see section \ref{detailsvollmer12}).

It is important to know if star formation occurs in the gas stripped from galaxies, and what its properties are, as this provides clues to the complex physics of star formation. The determination of constraints will allow us to progress in the modeling of the physics involved \citep[e.g.][]{kapferer09,tonnesen12}. It may also be crucial for understanding the nature of the Intra Cluster Light \citep[see e.g. the discussion in][]{tonnesen12} and of the stellar populations found in the halos of galaxies. 
Empirical studies are often based on a small number of galaxies and they provide few quantitative constraints on this kind of star formation. 
Our aim here is to complement these with a systematic study of large galaxies for which the occurrence of a recent episode of ram-pressure stripping has been clearly established, using an appropriate tracer of star formation. We focus on galaxies for which both \HI\ gas distribution maps and deep optical images are available, in order to reveal tidal disruptions (the present work is focused on possible star formation in ram-pressure stripped gas, while star formation in tidal tails will be studied in another GUViCS paper).

Recent star formation can be studied through a variety of tracers \citep[see e.g.][]{kennicutt98} such as the H$\alpha$ emission line.
The H$\alpha$ emission of Virgo galaxies has been frequently studied and discussed \citep[e.g.][]{kennicutt83,gavazzi02,koopmann04b,gavazzi06}.
However it may not be the best tracer for the low-level star formation we want to investigate in this work, due to the fact that it is not always related to in-situ star formation. For instance \citet{gavazzi01} discovered a 75 kpc-long H$\alpha$ tail behind two irregular galaxies in the Abell 1367 cluster which was associated with a radio continuum tail, but not with stellar emission. They suggested that the gas was first ionized inside the galaxy, and only then stripped. Another example is \object{NGC 4388} in the Virgo cluster, where the ram-pressure stripped gas observed out to 35 kpc from the galaxy disk was found to be ionized mostly by the radiation of an active nucleus rather than by star formation \citep{yoshida04}, even if a more recent study indicates that some low level of in-situ star formation cannot be excluded (Ferri\`ere et al., in preparation). Another example of an extra-galactic ionized region due to an AGN can be found in \citet{jozsa09}.

It has been shown that it is easier to study the star formation rate (SFR) in low-density gas using the far-ultraviolet (FUV) emission as a tracer. Indeed, FUV observations taken with GALEX at $\sim$1539 \AA, even typical one-orbit images, have revealed low levels of star formation in many cases, e.g. in \object{NGC 4438} \citep{boselli05}, \object{NGC 4262} \citep{bettoni10},\object{VCC 1217} \citep{fumagalli11}, \object{VCC 1249} \citep{arrigoni12} and more generally in the outermost parts of spirals and XUV (Extended Ultraviolet) galaxies \citep{gil05,thilker05,thilker07}.
In the FUV studies can be made of star formation at low gas densities and out to very large radii where the H$\alpha$ emission is more difficult to use \citep[e.g.][]{boissier07}, due to stochastic sampling of the IMF, observational difficulties \citep[see e.g.][]{goddard10}, or variations of the star formation history on short time-scales \citep{boselli09}. Moreover, the FUV emission traces star formation on time-scales of a few 100 Myr, consistent with the typical duration of stripping events (while H$\alpha$ traces star formation on a 10 Myr timescale), making it more appropriate for this work.
The exact star-formation history of a galaxy undergoing ram-pressure stripping might be complicated, which implies that the presently observed FUV emission may not be blindly converted into an SFR, i.e., by simply using conversion factors which assume a constant SFR. One solution is to try to make very detailed models of individual galaxies and to constrain them with a suite of multi-wavelength data \citep[see for instance the studies of \object{NGC 4330} and \object{NGC 4569} by respectively][]{vollmer12ngc4330,boselli06}. In our work, we adopt a phenomenological approach by measuring the FUV emission associated with ram-pressure stripped gas. In the case of a complicated gas stripping (and star formation) history, as may have occurred in some of our galaxies, the SFR will be poorly determined in a quantitative sense, but our derived constraints on the FUV emission (a record of star formation over the past 100 Myrs ) will still be useful.

GALEX FUV images and \HI\ maps have been used by \citet{bigiel10} to quantify the star formation law in low density gas in the outer regions of spirals and in dwarfs.
Similarly, we can use the deeper FUV images of the GALEX Ultraviolet Virgo Cluster Survey (GUViCS) \citep{boselli11} to constrain the level of star formation occurring in \HI\ gas that was removed from Virgo cluster galaxies by ram-pressure stripping.
In addition, we will compare the \HI\ distributions to the deep optical images of the Next Generation Virgo Survey (NGVS) \citep{laurasub} to reveal old stellar populations (which may indicate tidal origins), or the presence of stars that formed in the last few 100 Myrs).

In section \ref{sec:sample}, we present the sample of galaxies we selected for our study and the methodology adopted for our work. In section \ref{sec:comb}, we derive generic constraints on the star formation rate in the ram-pressure stripped gas in all galaxies from our sample, including a study of the ``Schmidt Law'' in the \HI\ tails. A discussion for each galaxy is given in the Appendix. A summary is presented in section \ref{sec:conclu}.

\section{Sample, data and method}
\label{sec:sample}
\subsection{A sample of ram-pressure stripped galaxies}

\begin{figure*}
\includegraphics[width=0.98\textwidth]{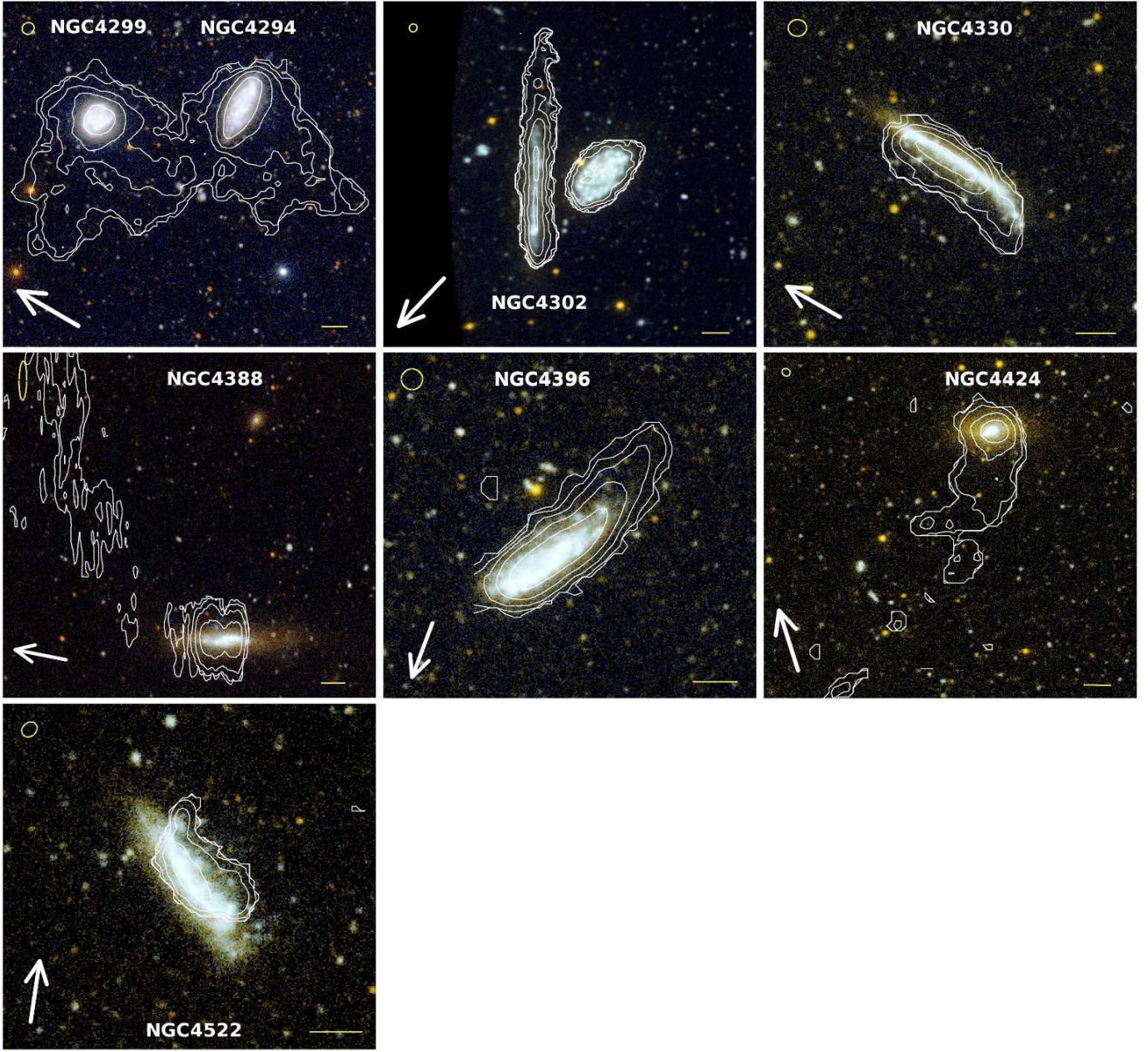} 
\caption{FUV/NUV images with \HI\ surface density contours superimposed. The contours shown correspond to \HI\ surface density levels of $log \Sigma_{HI} (M_{\odot} pc^{-2}) =-1, -0.5, 0, 0.5$ and $1$. The \HI\ beam is shown as an ellipse in the top-left corner of each panel. The horizontal bar in the bottom-right corner of each panel indicates a scale of 1 arcmin (or 4.9 kpc at the assumed cluster distance of 17 Mpc). The arrow indicates the direction of the cluster center. 
} \label{FigUVdata}
\end{figure*}

\begin{figure*}
\includegraphics[width=0.98\textwidth]{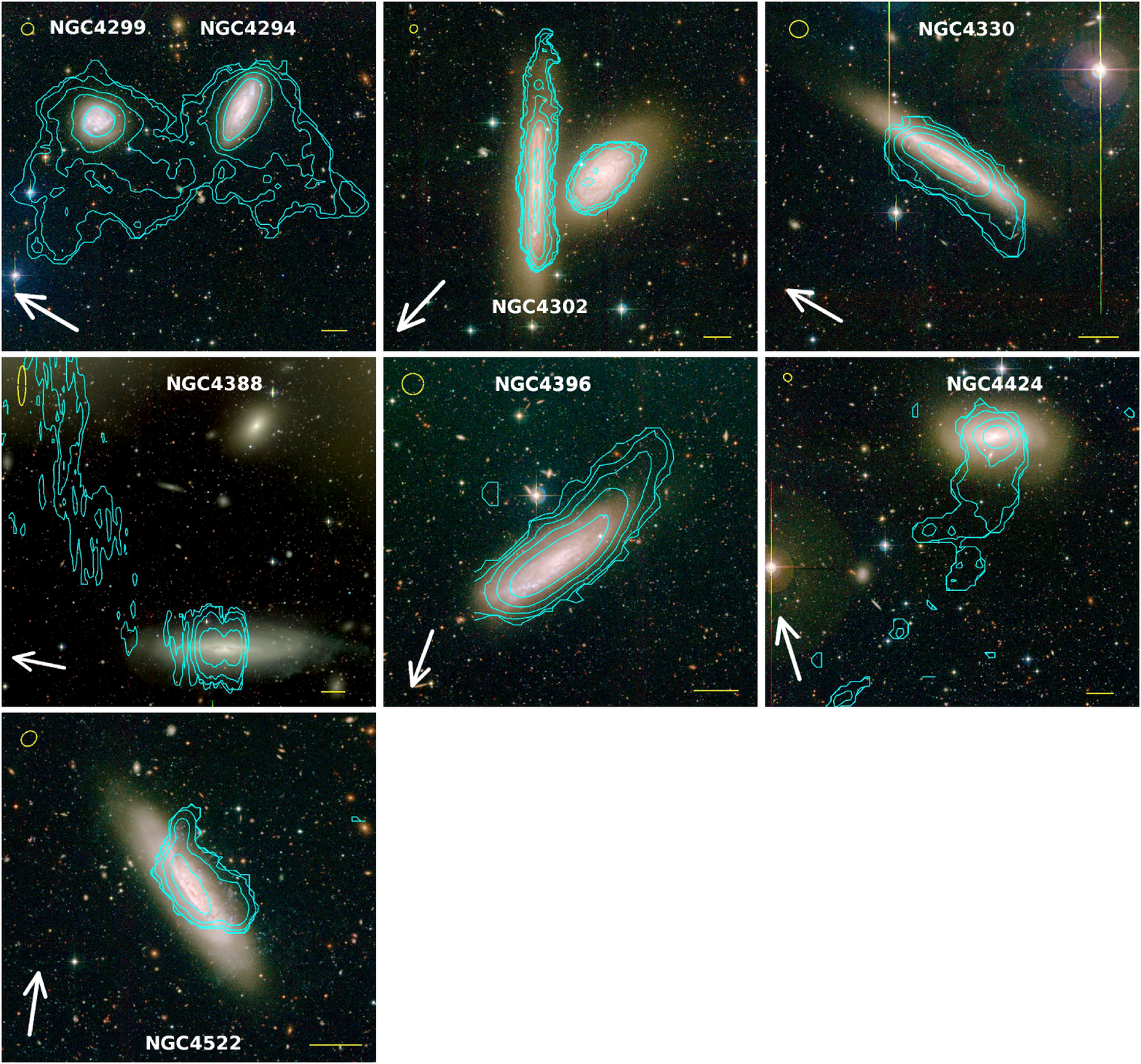} 
\caption{NGVS color images (where $u,g,i$-band data are represented as blue,green,red respectively), with \HI\ surface density contours superimposed. The contours shown correspond to \HI\ surface density levels of $log \Sigma_{HI} (M_{\odot} pc^{-2}) =-1, -0.5, 0, 0.5$ and $1$. The \HI\ beam is shown as an ellipse in the top-left corner of each panel. The horizontal bar in the lower-right corner of each panel indicates a scale of 1 arcmin. The arrow indicates the direction of the cluster center. }  \label{FigNGVSdata}
\end{figure*}

We aim at studying a sample of Virgo cluster galaxies i) which are likely to be undergoing a ram-pressure stripping event, ii) for which ram-pressure is likely to be the only (or main) form of interaction and iii) for which we have information on the \HI\ distribution.

\citet{chung07} analyzed the \HI\ distributions of seven galaxies with long one-sided \HI\ arms in the Virgo cluster, and suggested that these objects are very likely undergoing ram-pressure stripping events. Among these seven, six have FUV data available from the GUViCS compilation by \citet{boselli11}, with longer exposure times than the typical one-orbit, 1500 sec of the GALEX Nearby Galaxy Survey (NGS) allowing to reach a surface brightness limit of about 28.5 mag arcsec$^{-2}$. 

They are included in this study: \object{NGC 4294}, \object{NGC 4299}, \object{NGC 4302}, \object{NGC 4330}, \object{NGC 4396} and \object{NGC 4424}. The seventh galaxy, \object{NGC 4654}, only has FUV data from the GALEX All Sky Imaging Survey (AIS) which are too shallow for the purpose of this work. 
In addition, we included in our sample two other GUViCS survey galaxies which also have \HI\ tails due to ram-pressure stripping, as shown by previous studies: \object{NGC 4388} \citep{vollmer03} and \object{NGC 4522} \citep[e.g.][]{vollmer04,kenney04}.

Although other objects have been considered in studies of ram-pressure stripped tails \citep[e.g.][]{tonnesen12}, we chose not to include them here since they do not follow the criteria listed at the beginning of this section.

\subsection{Data}

VLA \HI\ maps from \citet{chung09} are used for all but one of the galaxies, with typical spatial resolutions between 15 and 30 arcsec (corresponding to 1.2-2.4 kpc at the adopted Virgo cluster distance of 17 Mpc), in which the gas external to the main bodies of the galaxies is well resolved. For \object{NGC 4388}, we use the more extended WSRT map of \citet{oosterloo05}. The beam size for each galaxy is listed in Table \ref{table:sample}.

GALEX FUV images, obtained at 1538.6 \AA{} effective wavelength with about 5 arcsec resolution \citep {morrissey07}, are taken from the GUViCS project \citep{boselli11}, a collection of archival and new GALEX fields covering a large fraction of the Virgo cluster area, which were all processed with the latest version of the GALEX pipeline (ops-v7, corresponding to GR6 products).
When several images overlapped at the position of a given galaxy, they were visually inspected to choose which to combine (using the task imcombine in IRAF\footnote{IRAF is distributed by the National Optical Astronomy Observatory, which is operated by the Association of Universities for Research in Astronomy (AURA) under cooperative agreement with the National Science Foundation.}) or to exclude (galaxy too close to a field edge or exposure too short). 
FUV images were used rather than NUV ones as the FUV emission is more closely related to recent star formation \citep[the FUV-NUV color evolves quickly with time after a star formation event, see e.g.][]{boissier08}. We actually inspected all NUV images but they do not reveal anything that is not already visible in the FUV, and they are therefore not used further in this paper (except for illustration in Fig. \ref{FigUVdata}). A qualitative discussion of the UV images of some of the galaxies was given in \citet{chung07}, and in a few previous studies of individual galaxies (see Appendix). Compared to these publications, we use the most recent data from the GUViCS project.
The resulting FUV images do not have a uniform exposure time (see Table \ref{table:sample}) but they are deeper (by factors from 1.3 up to more than 10) than the typical one-orbit Nearby Galaxy Survey (NGS) images. Combined FUV/NUV images are shown in Fig. \ref{FigUVdata}, with \HI\ contours overlayed, showing the extent of the \HI\ disks and tails.

Deep optical images of the galaxies in our sample were also made with Megacam on the CFHT in the $u$, $g$, $i$ and $z$-bands as part of the NGVS survey \citep{laurasub}. The exposure times are given in Table \ref{table:sample}. The NGVS images reach a 2 $\sigma$ limiting surface brightness of about 29 mag arcsec$^{-2}$ in the $g$-band, and they have been acquired with a seeing better than 1 arcsec \citep[the median FWHM is about 0.8 arcsec in the $g$-band,][]{laurasub}. Color composites of the NGVS images for our 8 objects are shown in Fig.\ref{FigNGVSdata}.

Finally, for all galaxies we also examined the H$\alpha$ images from the GOLD Mine database \citep{gavazzi03}, which are mainly taken from \citet{boselli02}. 
In general, in these images no clear H$\alpha$ emission is detected in the extended gas area (note that the field of view was too small to probe this area for \object{NGC 4294} and \object{NGC 4330}). For \object{NGC 4388}, an extended emission line region was found by \citet{yoshida02}, which is also seen in the H$\alpha$ image of \citet{kenney08}.

These H$\alpha$ images will not be used any further in this analysis because i) they come from a variety of sources and therefore have very different background quality and spatial extent, making it difficult to use them in a homogeneous way and ii) as mentioned in the introduction, the H$\alpha$ emission may well be a less effective star formation tracer than the FUV emission.

Basic properties (morphological type, B-band 25 mag arcsec$^{-2}$ isophotal radius $R_{25}$, axis ratio and major axis position angle from NED and GOLD Mine) of the selected galaxies are summarized in Table \ref{table:sample} together with exposure times. We have adopted a Virgo cluster distance of 17 Mpc \citep{gavazzi99} for all our galaxies \citep[see also][]{mei07}.

\begin{table*}
\caption{Sample}             
\label{table:sample}         
\centering                   
\begin{tabular}{l l l l l l l l l r l l }        
\hline\hline                 
Galaxy  & Type             & FUV exposure & $u$ & $g$ & $i$ & $z$   & 2 $\times$ $R_{25}$    & axis & $PA$ & \HI\ beam & pixel size   \\    
NGC        & (RC3/NED)        & (s)  &  (s)   &   (s)  &  (s)  &(s) & (arcmin) &  ratio     & (deg) & (arcsec) & (arcsec) \\ 
\hline                        
4294 & SB(s)cd          & 1689 + 3862          & 6402 & 3170 & 2055 & 4400 & 3.95  &  0.31  &  60 & 28.9 $\times$ 26.7 & 10 \\
4299 & SAB(s)dm?        & 1689 + 3862          & 6402 & 3170 & 2055 & 4400 & 1.96  &  1.00  &   0 & 28.9 $\times$ 26.7 & 10 \\
4302 & Sc? edge-on      & 18131                & 6402 & 3170 & 2055 & 3850 & 6.74  &  0.24  & 265 & 16.8 $\times$ 15.7 &  5 \\
4330 & Scd?             & 1689 + 3862          & 6402 & 3170 & 2055 & 3850 & 5.86  &  0.25  & 320 & 26.4 $\times$ 24.0 & 10 \\
4388 & SA(s)b? edge-on  & 1604 + 1590          & 6402 & 3170 & 2055 & 3850 & 5.10  &  0.24  &   0 & 95.1 $\times$ 18.0 &  9 \\
4396 & SAd? edge-on     & 2152 + 1026 + 2415   & 6402 & 3170 & 2055 & 4400 & 3.36  &  0.30  &  33 & 27.4 $\times$ 26.8 & 10 \\
4424 & SB(s)a?          & 2008                 & 6402 & 3170 & 2055 & 3850 & 4.33  &  0.50  &   0 & 17.6 $\times$ 15.5 & 10 \\
4522 & SB(s)cd? edge-on & 2496                 & 6402 & 3170 & 2055 & 4400 & 4.04  &  0.25  & 125 & 18.9 $\times$ 15.2 &  5 \\
\hline                                  
\end{tabular}
\end{table*}

\subsection{Methodology}

In the remainder of this paper, we will focus mainly on the comparison between the \HI\ distribution and the GALEX FUV emission. In regions of low gas density the molecular fraction is low \citep[e.g.][]{leroy08}, which makes the \HI\ a good tracer of the total gas. Furthermore, the FUV emission is a well known indicator of the star formation rate \citep[e.g.][]{kennicutt98}. In order to be able to compare our results with the work of \citet{bigiel10}, who studied dwarfs and outer regions of spiral galaxies, we will use their conversion from FUV intensity, $I_{FUV}$, to SFR surface density, $\Sigma_{SFR}$ (assuming a Kroupa-type IMF):
\begin{equation}
\label{eq:calib}
\rm \Sigma_{SFR} [M_{\odot} \, yr^{-1} \, kpc^{-2}] = 0.68 \, 10^{-28} \times I_{FUV} [erg \, s^{-1} \, Hz^{-1}\, kpc^{-2}].
\end{equation}
This conversion assumes a constant SFR over the time-scale of the UV emission (about 10$^8$ yr), which may not be the case at low densities. For this reason, most of our figures show the observed surface brightness as well as the converted SFR.

A potential limitation is the possible attenuation due to dust. Its effect can be corrected when a dust tracer is available (e.g. far-infrared emission). 
While we can get an idea of the global dust attenuation for our galaxies from \citet{cortese08}, we cannot know its detailed distribution in each object, especially in the extended \HI\ tails analyzed in this work.
Still, dust extinction is known to decrease with radius within galaxies \citep[e.g.][]{boissier04,boissier07,holwerda05,munos09}, and it is also modest in Low Surface Brightness galaxies \citep{hinz07,rahman07,boissier08}. Since this study is focused on the outermost, low-density regions of galaxies, it is likely that the dust attenuation, if present, is small. 
%
On the other hand, there are indications that dust stripping occurs together with gas stripping \citep{cortese10, cortese12}: both the gas and dust disks of gas-deficient cluster galaxies are truncated (though the dust is less affected than the \HI\ gas). Also, Ferri\`ere et al. (in preparation) note dust lanes seen in absorption coinciding with the \HI\ peaks in the tail coming from \object{NGC 4388}.
However, no far-infrared emission from dust has been observed in stripped tails, suggesting again small amounts of dust and low attenuation.
As a result, the FUV emission should be a direct indicator of the star formation rate (using equation \ref{eq:calib}). By comparing the \HI\ and FUV emission, we are thus to the first order comparing the amount of gas and the star formation rate, i.e., what is usually called the ``star formation law'' or Schmidt Law, after \citet{schmidt59}.

In the optical, the higher spatial resolution of the NGVS data allows us to easily recognize background objects or stars. The optical images also provide complementary information on the studied galaxies. For example, the observation of a relatively old stellar population (red colors) would suggest tidal stripping (as is occurring in e.g. \object{NGC 4438}) rather than ram-pressure stripping. On the other hand, if significant star formation occurred during the last 100-500 Myr (a slightly longer time-scale than the 100 Myr probed by the FUV emission) blue patchy emission should be seen at its location in the optical images.

Two methods were applied to study the galaxies:

1) Inspection of the FUV and optical images by eye. We attempted to detect any emission that may be associated with the \HI\ tails and the peaks therein -- it should be easy to detect young star clusters in such images. 
Although in their galaxies, \citet{chung07} did not find optical emission associated with the \HI\ tails on SDSS images, the much deeper NGVS images allow us to revisit this issue.

2) A pixel-by-pixel analysis. If star formation is relatively low-level and diffuse, it may not be so easily recognizable by eye, but a quantitative measurement of the FUV emission may well be possible. Star formation was indeed detected in the surface photometry of \citet{boissier07} and \citet{bigiel10} at low gas densities, typical of the outer parts of spiral galaxies. The method is described below.
\label{galexpsf}
We first performed a visual inspection of the FUV images to mask any objects in the foreground (stars) or background (galaxies), using the high resolution of the NGVS images to clearly identify them as such. The size of each mask was chosen by eye to cover the object, at the GALEX resolution.
Pixels within the masked areas were then replaced by an interpolation of the surrounding pixels, to prevent that the convolution done at a later stage would spread their flux outside the mask. The resulting images were inspected to make sure that no obvious structure was left after this interpolation.
Note that the GALEX FUV PSF has a Gaussian core of 5 arcsec width, but also extended exponential wings at a level of $\sim$ 10$^{-4.5}$ of peak value at 60 arcsec radius (see the GALEX technical documentation, chapter 5\footnote{http://www.galex.caltech.edu/researcher/techdocs.html}).
While we made sure that the area corresponding to the PSF core is masked, it is possible that some of the flux of background and foreground objects is spread out to large distances due to the wings, which we cannot mask systematically. We may therefore slightly overestimate the UV emission level at their position and around them. However, most of the objects are small or faint (background galaxies). In any case, foreground and background objects have a spatial distribution independent from that of the \HI\ tail. This effect thus only contributes to the FUV sky noise level.
The resulting FUV images were then convolved with a Gaussian kernel in order to approximately reproduce the \HI\ beam size for each galaxy. They were then re-sampled at the same pixel size as the \HI\ maps (5 or 10 arcsec, depending on the galaxy, see Table \ref{table:sample}). The pixel-by-pixel studies were performed after further masking of bad image areas (pixels outside of the GALEX circular field of view and other artifacts) and other galaxies.
In the case of \object{NGC 4388}, because the original \HI\ map uses rectangular pixels due to its elongated beam shape, we re-sampled both the FUV and \HI\ images with square pixels of 9$\times$9 arcsec. 
The relatively large pixel sizes used in this study have the advantage to increase the signal-to-noise ratio per pixel in the FUV, which is useful at the low surface brightness levels studied, but they are still small enough to allow the determination of the shapes of the galaxies and their tails.
The fact that the pixels are smaller than the beam size (our working resolution) implies that our pixels are not independent. However this will not affect our study, in particular when we compute averages within \HI\ density bins: increasing the pixel size is approximately equivalent to adding more pixels to the same \HI\ bin. We tested the effect of changing the pixel size from 10 to 30 arcsec on \object{NGC 4294}. After dividing the number of individual points by the scaling factor of 9, the binned results were very similar.
\begin{figure*}
\includegraphics[width=0.44\textwidth]{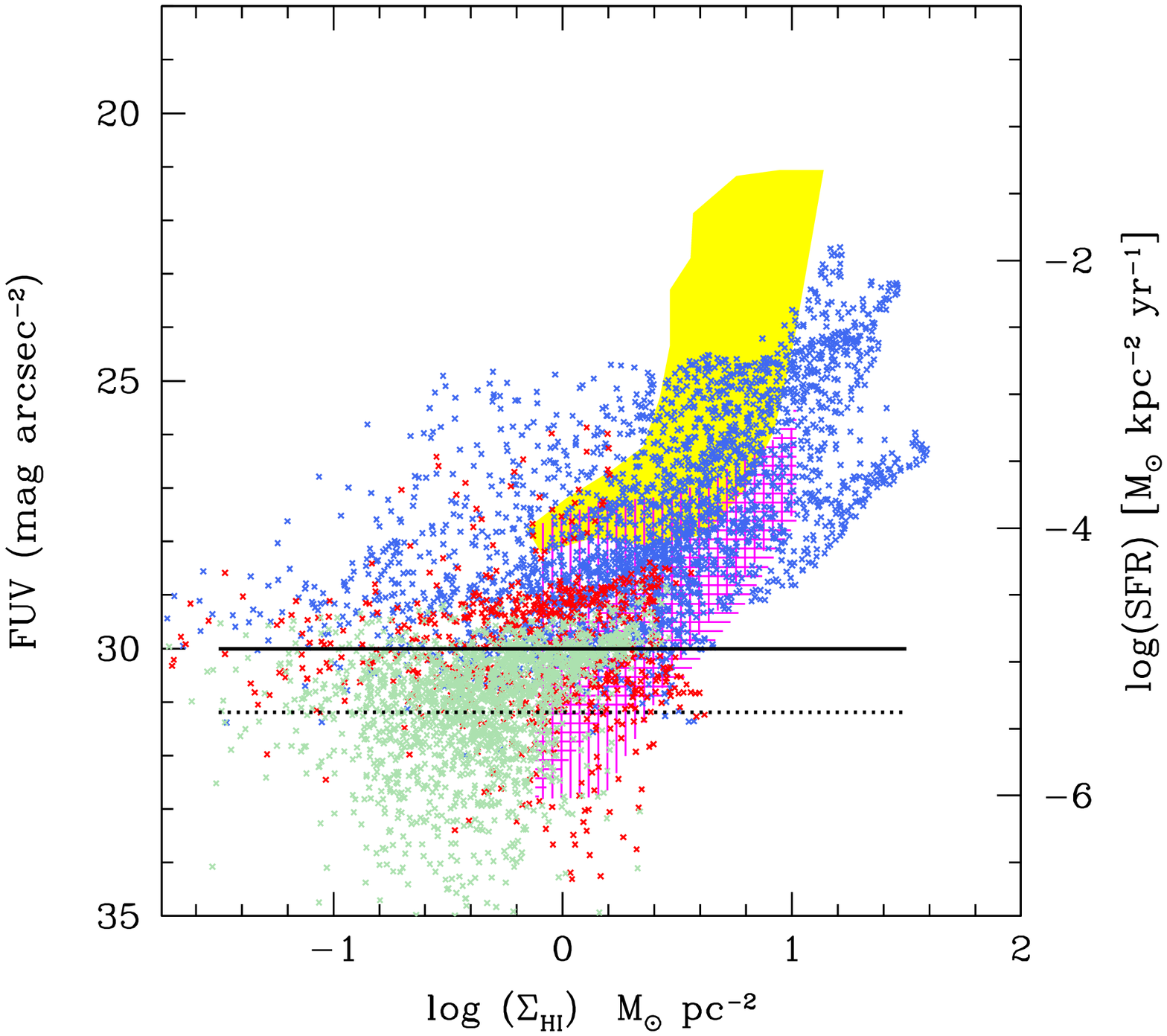}
\includegraphics[width=0.44\textwidth]{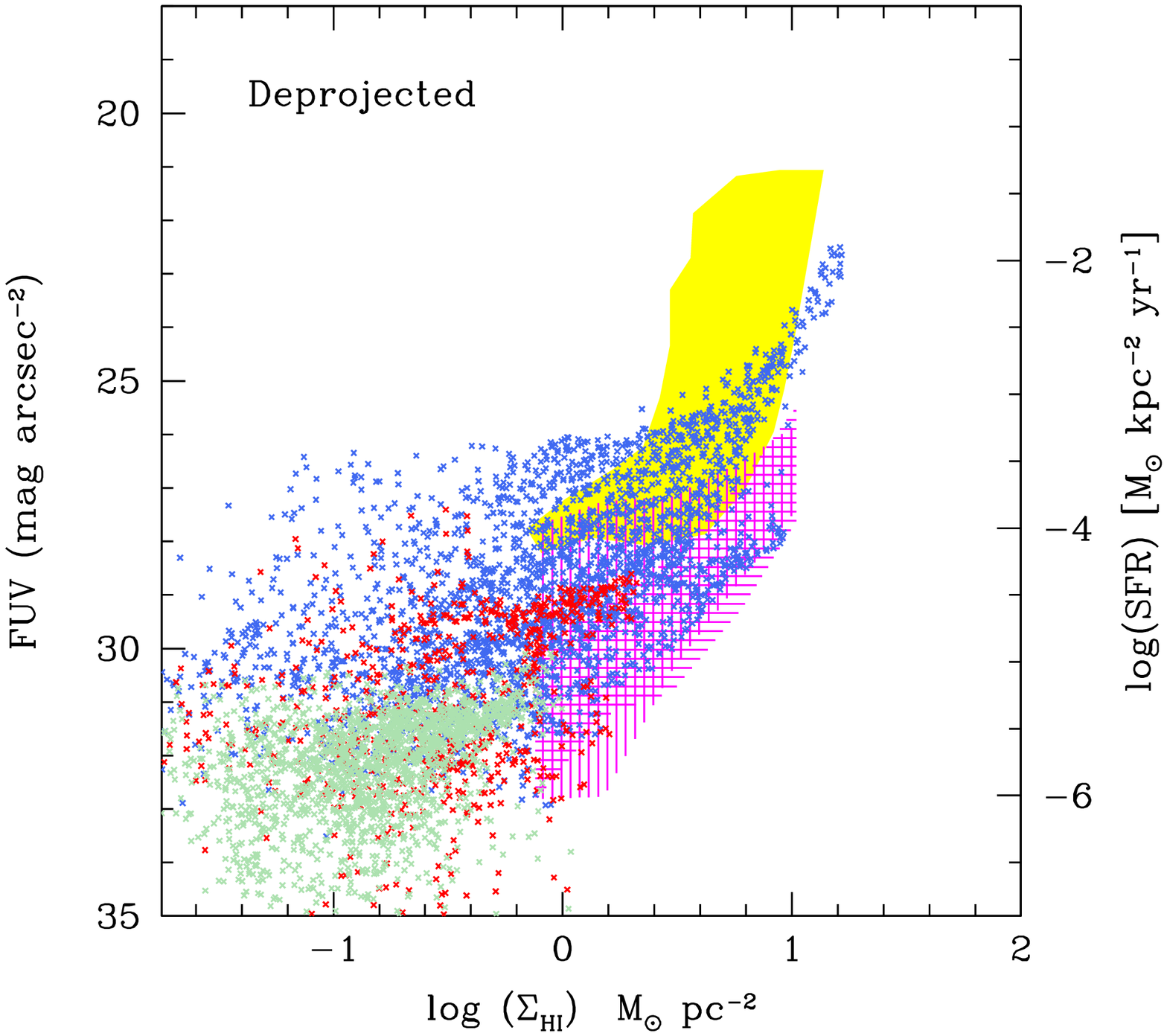} \\
\includegraphics[width=0.44\textwidth]{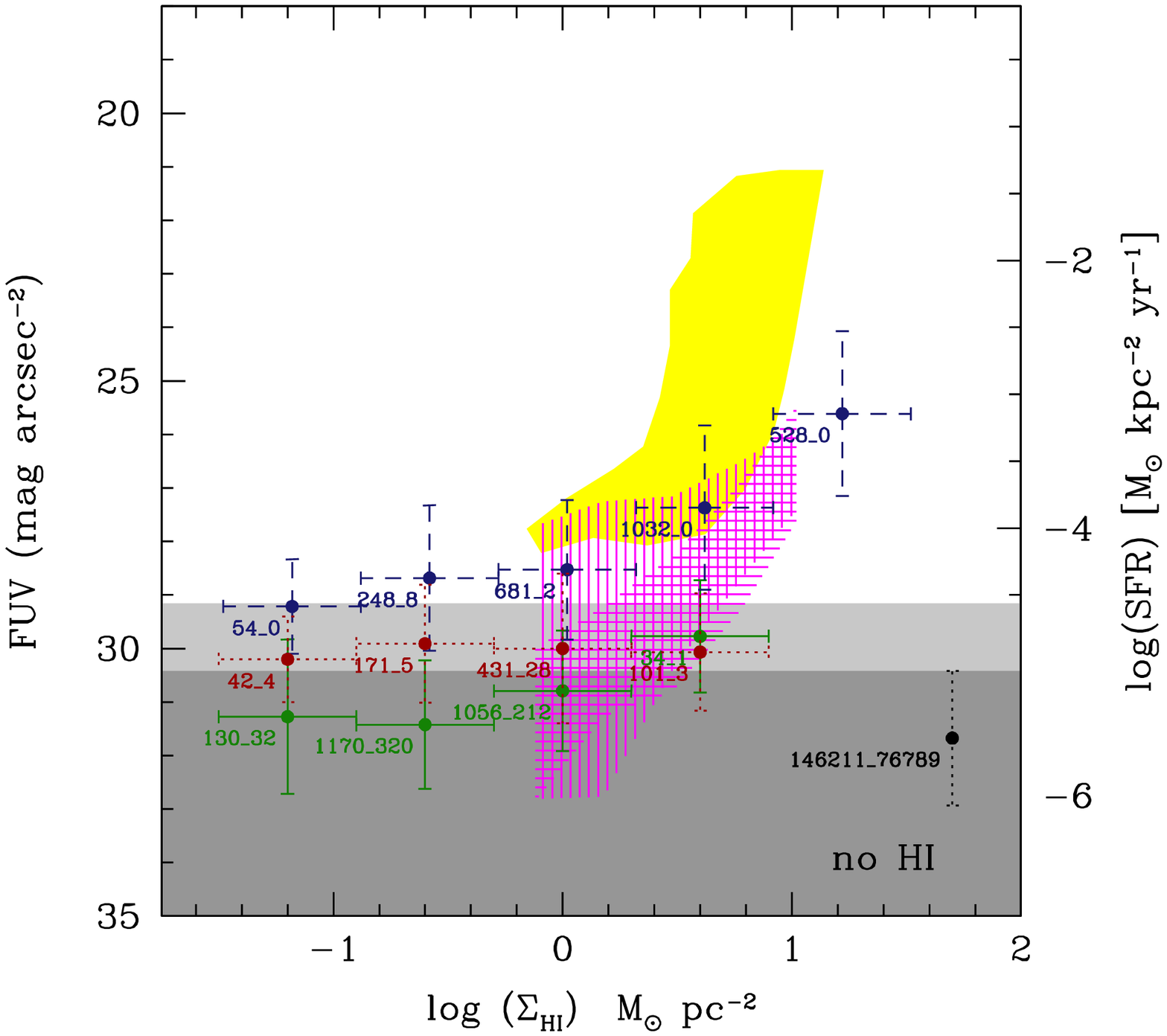}
\includegraphics[width=0.44\textwidth]{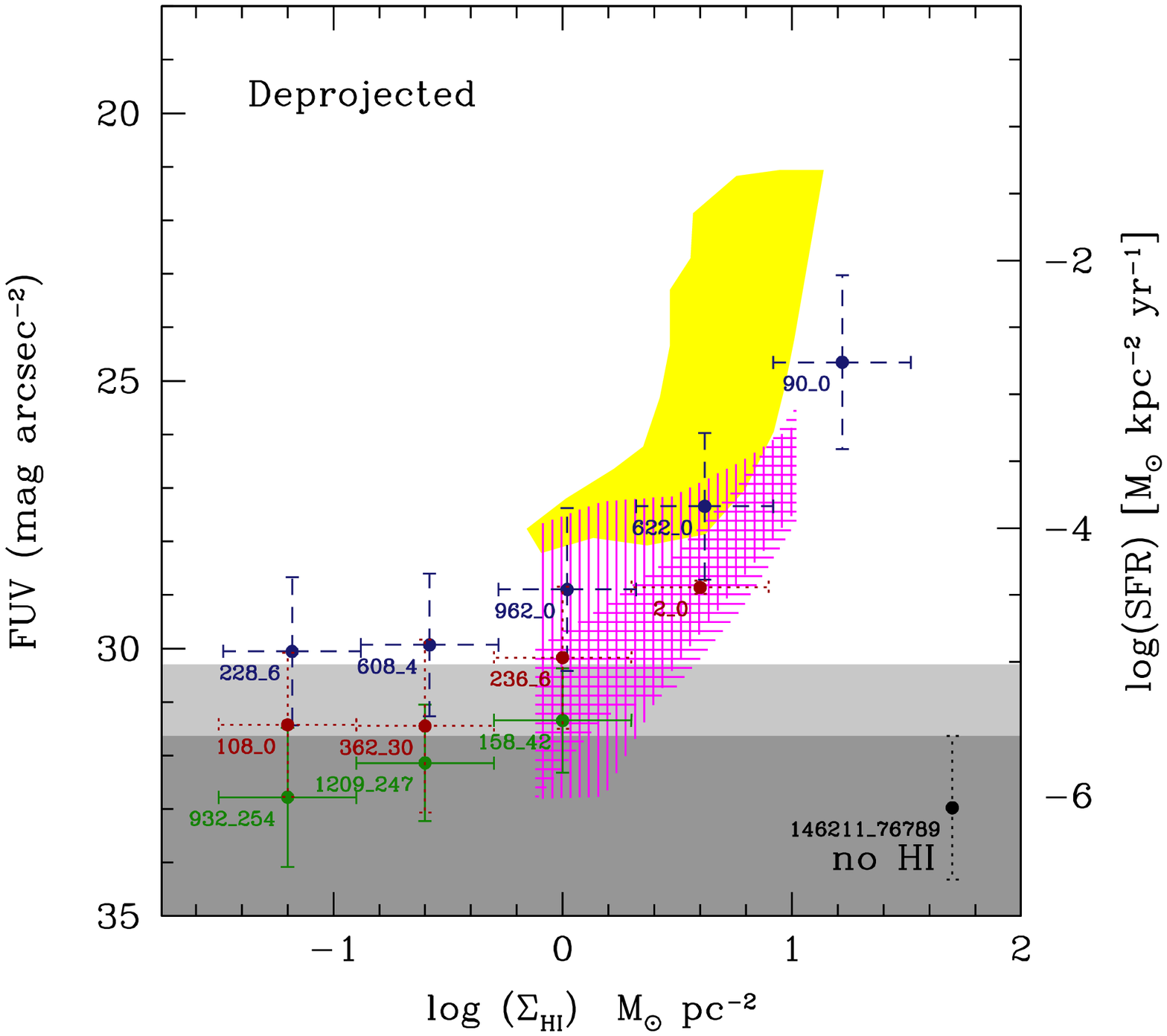} \\
\caption{The relationship between FUV surface brightness and \HI\ surface density, $\Sigma_{HI}$, in our full sample. The scale on the left shows surface brightness (in mag arcsec$ ^{-2}$), while the scale on the right shows log(SFR) (in M$_{\odot}$ yr$^{-1}$ kpc$^{-2}$), assuming the same FUV-SFR conversion as in \citet{bigiel10}. Top row: Blue/red/green points indicate pixels within, respectively, the inner galaxy, the extended galaxy and the external gas (see section \ref{sec:tagging} for the definition of these regions).
The typical 1 (3) $\sigma$ limit in FUV surface brightness is indicated by the horizontal dotted (solid) line.
Bottom row: The points with the error bars indicate, from top to bottom, the mean and its dispersion within the \HI\ surface density bins for each of the three regions (inner galaxy, extended galaxy and external gas). The two numbers on the left of each error bar, separated by an underscore symbol, indicate (left) the number of pixels used for this point, and (right) the number of pixels with a flux below the mean sky level in the bin. As using the latter pixels would give an infinite surface brightness they are not included in the computation of mean surface brightness. A large value for this second number is a good indication that we are very close to the sky value. The error bar in the lower-right corner (labeled ``no \HI'') shows the result of the same analysis when performed on the FUV emission outside the detected \HI\ only, which gives an indication of the detection limit with this method (the dark- and light-grey shaded area show, respectively, the corresponding 1 and 2 $\sigma$ limits).
The yellow-shaded area indicates the location of the nearby field galaxies from the THINGS sample, from \citet{bigiel08}. The vertically and horizontally shaded areas correspond respectively to dwarf galaxies and the outer parts of spirals (i.e., the points with error bars shown in Fig. 8 of \citet{bigiel10}).
The two left-hand panels correspond to the directly observed surface brightnesses and surface densities, while in the right-hand panels a geometrical correction was applied for the inclination of each galaxy. Similar figures are shown in the Appendix for each of our galaxies.
}         
\label{FigSLcomb}
\end{figure*}

\label{sec:tagging}
Since our goal is to compare the star formation law (the relation between  star formation rate and gas density) in the gas removed from galaxies via ram-pressure stripping with the law found within galaxies (and their outer parts), we tagged each pixel according to which of the following three regions it belongs: 
i) The inner galaxy, within an ellipse of semi major axis radius $R_{25}$, with inclination (axis ratio) and position angle ($PA$) as listed in Table \ref{table:sample}. 
This area is grown by $n$ pixels to avoid counting as ``extended'' the fraction of the flux originating from the inner galaxy which is spread out beyond its limit due to the resolution. This number $n$ is fixed as 30 divided by the pixel size (given in Table \ref{table:sample}). This is motivated by the fact that the resolution for most of our galaxies is somewhat smaller than, but close to, 30 arcsec, thus ensuring that the area is extended by at least 1 resolution element. In the case of \object{NGC 4388}, however, due to the very elongated \HI\ beam shape, this is true only along the major axis of the galaxy. Along the minor axis, and especially close to the center, some flux from the galaxy is spread out into the ``external'' region. We masked the area that may be affected by this effect outside the galaxy, and it is not taken into consideration in the following.
ii) An ``extended'' galaxy area, contained between the inner galaxy and a similar ellipse with a semi major axis radius of $1.5 \times R_{25}$. This area is grown by the same number of pixels as the previous one. 
iii) An ``external'' area, containing the remaining pixels outside the extended galaxy area. The three regions are shown in the right-hand part of each figure in the Appendix.

This division into three regions allows for each galaxy a comparison of the \HI\ and UV emission found inside and outside the galaxy, and a direct comparison of the UV-\HI\ relationship in various galactic environments.
In our sample, 15$''$ corresponds to 1.2 kpc. With our resolution (Table \ref{table:sample}), these three environmental regions are thus distinguished on a scale of about 1 to 2 kpc.

\section{A global picture of star formation in ram-pressure stripped gas}
\label{sec:comb}

\subsection{Summary of the behavior in individual galaxies}

The details of the study of each galaxy are provided in the Appendix. In two cases we clearly see FUV emission in the ram-pressure stripped \HI\ gas: \object{NGC 4330} and \object{NGC 4522}. The NGVS images also reveal blue clusters in these two galaxies and other structures that may be related to recent star formation and which call for further investigations.
In these two cases the gas is still close to the galaxy, however, and the FUV emission does not extend over the whole \HI\ tail. Our approach is more meaningful on these larger scales. 

In the other galaxies we see almost no FUV emission or NGVS blue clusters in the external gas, especially where the \HI\ is well extended and well separated from the galaxy. The pixel-by-pixel analyses suggest that the star formation in the external gas, if present, is very limited, and certainly lower than within the galaxy (or the extended galaxy region) even at the same gas surface densities. In most cases, the FUV emission in the tail is very close to the detection limit, and even consistent with zero star formation, despite \HI\ surface densities of up to 1 M$_{\odot}$ pc$^{-2}$.

Finally, mostly in \object{NGC 4522}, the level of the FUV emission inside the galaxy seems to be slightly higher than what is usually found in spirals \citep[e.g.][]{bigiel08} at the same gas surface density levels, at least for part of the pixels, despite the fact that we did not correct for extinction (and our SFR can thus be considered as a lower limit). This suggests that a very mild enhancement of the star formation efficiency \emph{inside} the galaxy during the interaction may be possible, in agreement with the H$\alpha$ study of \citet{koopmann04b}. However, as this enhancement is limited in both time (during the interaction) and space (in the inner part of the galaxy only), it is likely to play a small role in the global evolution of the galaxies \citep[see also][]{tonnesen12}, and  therefore not noticeable when considering integrated quantities \citep[e.g.][]{iglesias04} or large samples of galaxies as a whole. Indeed, the UV properties of the Herschel Reference Survey galaxies show no evidence for an important enhancement of star formation in their centers \citep{cortese12hrsuv}.

\citet{vollmer12sfe} recently made a study of the influence of the cluster environment on the star formation efficiency in 12 Virgo cluster galaxies. Their sample is defined using criteria very different from ours (multi-wavelength observations including Spitzer, GALEX, \HI\ and CO data). It includes galaxies undergoing various types of interactions, with two cases of ram-pressure stripping in common with our study: \object{NGC 4330} and \object{NGC 4522}. They used a pixel-based approach very similar to the one adopted here. Our approaches are complementary: they studied the relation between the star formation rate and the molecular (or total) gas distribution (also within the galaxies themselves), whereas we use only the neutral gas and focus on the outer regions, where no molecular gas is observed, and the \HI\ alone can be used to measure the global gas reservoir for star formation.
Our results are qualitatively consistent with theirs. They found that the molecular fraction (and star formation) is modestly increased within galaxies undergoing ram pressure, and that the star formation efficiency (SFR divided by the molecular or total gas amount) tends to be lower in extra-planar gas associated with ram-pressure stripping. 
\label{detailsvollmer12}

\subsection{The Star Formation Law in the stripped gas}
\label{sec:SL}

Figure \ref{FigSLcomb} shows the correlation between FUV surface brightness and \HI\ surface density at the pixel scale for our full sample. This figure allows us to try to find a general trend among the galaxies, with better statistics than when examining individually each of the figures for our galaxies (shown in on-line figures \ref{FigLocalN4294} to \ref{FigLocalN4522}, where the 3 and 1 $\sigma$ sky level has been indicated for each galaxy). Note however that most of the pixels contributing to the measurement of the star formation rate within \HI\ tails come from three galaxies only, namely \object{NGC 4294}, \object{NGC 4299} and \object{NGC 4388}.
The common \HI\ tail between \object{NGC 4294} and \object{NGC 4299} is of course counted only once for this combined figure.
As discussed in the Appendix, even barely detected pixels have been used to compute averages in \HI\ surface density bins, and the point with the error bar in the lower-right corner of each panel provides an estimate for the FUV emission level detection limit of our method.
In the left-row panels, we did not apply any inclination correction, whereas in those on the right a simple geometrical de-projection has been applied, multiplying the densities by the axial ratio determined at the 25 mag arcsec$^{-2}$ level (listed in Table 1). While within the galaxy it is reasonable to de-project in this manner (although with large uncertainties in edge-on galaxies), within the stripped gas inclination-corrected values for \HI\ column density and FUV surface brightness are harder to determine since the 3D geometry of the stripped gas is unknown. Nevertheless, the sets of observed and deprojected values provide two test cases. In the case of \object{NGC 4294}, the geometry of the galaxy and tail suggests that the galaxy feels the ram-pressure wind edge-on so that the deprojection may give reasonable estimates. On the other hand, geometries like those of \object{NGC 4388} or \object{NGC 4522} suggest a face-on wind, which makes a deprojection of the tail using the geometrical parameters of the galaxy irrelevant. In such cases, it is better to consider the directly observed quantities 

In this figure, the position of our pixels (and binned values) is compared to two important studies. First, \citet{bigiel08} established a new benchmark in the studies of star formation in nearby galaxies through combining all the observations required to study the Schmidt Law (the SFR-gas relationship) on a pixel-basis. Although their main conclusion is that the SFR correlates better with the molecular gas, their SFR-\HI\ relationship as shown in Fig. \ref{FigSLcomb} is representative of ``normal'' field galaxies. Moreover, when the highest surface densities are avoided, the \HI\ is a good proxy for the total gas surface density, as already discussed.
Second, \citet{bigiel10} extended this work by applying a pixel-by-pixel approach similar to ours to the outer part of spirals and to dwarf galaxies, both environments where the gas is likely to be dominated by \HI. The Schmidt Law they derived is also indicated in Fig. \ref{FigSLcomb} and it shows that the Star Formation Efficiency (SFE, defined as the SFR divided by the \HI\ mass per pixel) in the outer part of spirals and in dwarfs is smaller than within the inner spiral disks.

Our pixel-by-pixel approach is very similar to these two studies, which allows us to directly compare our results to theirs. Our SFR values at a given \HI\ surface density should lie within, respectively, the range of values measured by \citet{bigiel08} for our inner galaxies and those of \citet{bigiel10} for our extended rings.
Inside the galaxies, the relationship between the \HI\ surface density and the SFR is indeed found to be consistent with that found by \citet{bigiel08}, taking into account the fact that we did not correct for extinction so that our inner galaxy SFR is underestimated, especially for the largest \HI\ column densities.
We also find that at a given gas surface density level, the SFR is higher in the inner galaxy than in the outer disk, as expected.

The possible star formation in the tails can be directly compared with the findings in outer spirals and dwarfs from \citet{bigiel10}, or with our results in the ``extended galaxy region''. At a given gas density, the FUV emission (and thus the SFR) is lower in the \HI\ tails than in our extended galaxy regions or even in the outer, low density parts of normal galaxies, indicating a very low star formation efficiency. 
Since our tail pixels by definition lie beyond 1.5 $R_{25}$, our results would also be consistent with a decrease of the Star Formation Efficiency (the SFR/\HI\ ratio) with radius (even if attributing a galactocentric radius to the gas in the tails may be unphysical in some cases, given 
 the complex geometry of the tails).
The SFR measured at the $\Sigma_{HI} \sim 1$ M$_{\odot}$ yr$^{-1}$ level in the stripped-tail pixels is lower that the SFR measured in our external galaxy regions. It is marginally consistent with the broadly dispersed values from \citet{bigiel10} for dwarfs and outer parts of spirals, but clearly on the low side. Finally, the SFR in the tails is about 10 times lower than within the galaxies. This level is quite close to our detection threshold however, and could even be considered an upper limit.

In fact, the FUV emission level in the gas tails is always consistent with being zero, except in the highest \HI\ surface density bin, where it is barely above the average value found in pixels without any \HI. This FUV emission never exceeds the levels found in the inner galaxies or in the external regions. The images show that the FUV emission in the tails is in general not associated with \HI\ peaks far from the main body of the galaxies, but rather that it occurs mostly close to their outer edges. In order to smooth from the UV to the \HI\ resolution we convolved the UV images with a Gaussian kernel, but it should be noted that neither the \HI\ beam nor the FUV PSF are perfectly gaussian or absolutely symmetric. The results of this transformation could thus introduce imperfections, including some pollution at the edge of the galaxies coming from the UV light of the galaxy itself.
We already mentioned the extended exponential wings of the GALEX PSF in section \ref{galexpsf}.
We checked that the FUV emission level at faint gas column densities is not solely due to these wings (contrary to the case of foreground and background objects, the relative positions of the \HI\ tail and the galaxy are not independent from each other). This was done by comparing two convolutions of the bright parts of the galaxies \object{NGC 4294}/\object{NGC 4299} (at $\leq$10 $\sigma$ above sky level): one made with the observed PSF, and another using a modeled PSF with extrapolated exponential wings extending out to 6 arcmin radius. We found that the level of emission in the external pixels that could be attributed to the smearing out of the bright part of the galaxy by the PSF wings is at most a few percent of the measured level.

In any case, Fig. \ref{FigSLcomb} puts strong constraints on the SFR in the outer gas tails: it is much lower than within galaxies, at the same gas surface density level -- at least one order of magnitude at about 1 M$_{\odot}$ yr$^{-1}$. 
These quantitative constraints should be taken into account in future physical models of ram-pressure stripped gas that include star formation \citep[e.g. in the line of the work of][]{tonnesen12}.

\subsection{Constraints on the amount of star formation outside the galaxies}
\label{sec:frac}

\begin{figure}
\includegraphics[width=0.5\textwidth]{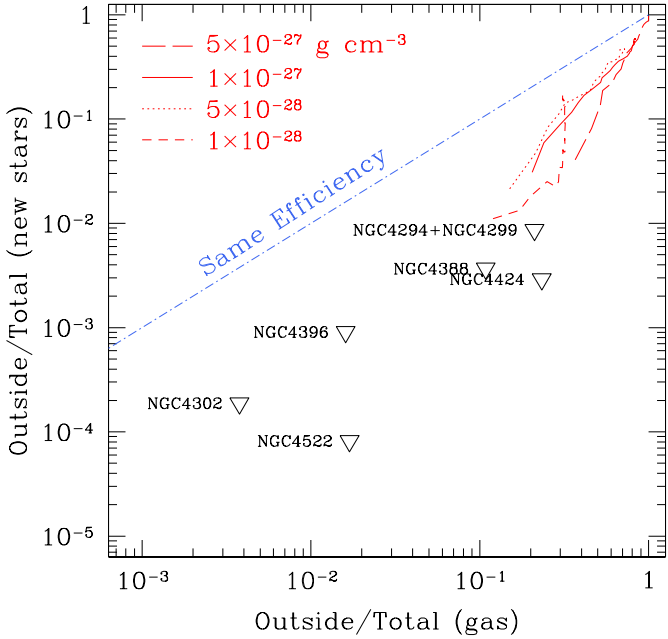}
\caption{Mass fraction of newly formed stars outside the galaxy as a function of the fraction of \HI\ gas outside the galaxy. The blue line indicates the 1:1 relationship which would be followed if star formation had the same efficiency inside and outside of galaxies. The red curves show the evolution of the \citet{kapferer09} models for a relative velocity of 1000 km s$^{-1}$ between the galaxies and the ICM (appropriate for the Virgo cluster), for different surrounding densities, as indicated in the top-left corner \citep[for comparison, the average Virgo cluster density is 3 $\times$ 10$^{-27}$ g cm$^{-3}$, see][]{bosellirev}. The triangles show the same fractions computed for the galaxies in our sample, using the FUV light (corrected for internal extinction, see section \ref{sec:frac}) and the \HI\ mass (assuming it traces the total gas mass). }  
\label{FigFractions}
\end{figure}

\citet{kapferer09} produced models for star formation in galaxies undergoing ram-pressure stripping and in their tails. They show the evolution over a period of 400 Myr of the mass fraction of newly formed stars in the tail with respect to the entire galaxy, as well as the fraction of gas in the tail.
In Fig. \ref{FigFractions} we compare the evolution of these mass fractions (showing them only for values exceeding 1 percent). The models evolve with time from the lower-left end of the curves (when most of the gas is still inside the galaxy) up towards the right (once the gas has been pushed outside the galaxy), with typically a rapid evolution in the first $\sim$ 100 Myr. 
This figure provides an interesting diagnostic of the efficiency of star formation in the tails with respect to that inside the galaxies. Indeed, if the efficiency were similar in these two environments, galaxies should lie on the 1:1 relationship indicated in the figure. The models lie below this curve, however: the star formation efficiency is lower in the external gas than inside the galaxy.
\citet{tonnesen12} also simulated star formation in ram-pressure stripped galaxies. They find a lower level of star formation in the tail than \citet{kapferer09}: only 1\% of their newly formed stars are in the tail after 250 Myr.

In the same figure, we show the positions of our galaxies, assuming that the FUV tracks the young stars and the \HI\ traces the total gas mass, and integrating the fluxes and masses in the tails, outside of the 1.5 $R_{25}$ radii of the galaxies' outer regions. 
The reader should take into account the fact that \object{NGC 4302}, \object{NGC 4396} and \object{NGC 4522} have very few pixels well outside the galaxy according to our definition of the three regions, explaining the very small fraction of external gas and SFR and making it hard to compare with the models.

We applied a dust attenuation correction to the flux from the inner galaxy, based on the global attenuation from \citet{cortese08} (median 1.3 magnitude, ranging from 0.8 to 2.3). Contrary to the \emph{local} analysis of the Schmidt Law in the \HI\ tails of section \ref{sec:SL}, such a correction is now necessary since we compute \emph{global} fractions depending on the inner-galaxy UV fluxes, which are affected by dust extinction.

Our points are clearly situated in the lower part of the diagram, pointing to a low efficiency of star formation outside the galaxies. The three objects with relatively large fractions of external gas (in the right-hand part of the figure) can be compared to the models. They seem to indicate lower levels of star formation than predicted by the models for a similar external gas fraction.

The comparison is still rather qualitative because our study uses observed FUV and \HI\ data while the simulations make predictions in terms of newly formed stars and gas. 
The models represent also an idealized case, whereas our sample includes a mix of e.g. geometry, relative velocity and ages. For this reason the regions considered as being ``outside'' and ``inside'' a galaxy are not determined rigorously in the same manner. In the models the separation is simply made at 10 kpc above the disk plane, while in the observations we make the separation at a specific optical isophotal radius of the galaxy.
Nevertheless, our diagram can serve as a constraint for future models that will include detailed physics.

\begin{table}
\caption[]{Derived properties of the tails.}
\label{tabtailq}
\begin{tabular}{l l l l l }
\hline
System         &  $log(SFR)$         & $log(M_*)$   &$<\mu_{FUV}>$ &  $log(\Sigma_*)$ \\
(NGC)          & (M$_{\odot}$ yr$^{-1}$) & (M$_{\odot}$) & (mag arcsec$^{-2}$) & (M$_{\odot}$ kpc$^{-2}$)\\
\hline
4294+4299      &  -2.1       & 5.7    &    30.6     &  2.7\\
4302           &  -5.0       & 2.8    &    31.9     &  2.2\\ 
4388           &  -2.8       &  5.1   &    32.2     &  2.1\\
     4396      &  -3.6       & 4.2    &    31.1     &  2.5\\
     4424      &  -3.4       & 4.5    &    32.2     &  2.1\\
     4522      &  -4.8       & 3.0    &    32.6     &  1.9\\
\hline
\end{tabular}
\end{table}

The models of \citet{tonnesen12} show the evolution of the SFR in the gas tail and of the stellar mass created in the tail during the ram-pressure stripping event (their Fig. 14), as well as the range of stellar surface densities in the tail (their Fig. 13). 
From our data, it is possible to make estimates of similar quantities, although it is not completely straightforward. To compute the integrated 
SFR and an average surface brightness in the tail, we first have to assume a conversion from FUV flux to SFR (see equation 1). Then we have to integrate over the pixels that lie within the \HI\ tails, outside the galaxies. To compute stellar masses and stellar densities, we have to make a further assumption. We estimate that the SFR has been about constant for the last 100 Myr, with a mass fraction $R$ returned to the interstellar medium \citep[see e.g.][]{prantzos09} of $R \sim$ 0.3. The total stellar mass in the tail (in solar masses) for a given SFR is thus simply 
SFR (M$_{\odot}$ yr$^{-1}$) $\times$ 10$^8$ $\times$ (1-$R$).
Note that our stellar mass density estimates are derived from FUV data while it is customary to use longer wavelengths when estimating stellar masses. However, in this case, we only consider the \emph{young} stars created over the last few 100 Myrs in the \citet{tonnesen12} models, to which the FUV is more sensitive.
The results of these computations are given in Table \ref{tabtailq}, and we compare them below to the data given in \citet{tonnesen12}. Again, the comparison is not perfect because we can only estimate orders of magnitude with our assumptions, and because of the large differences between the galaxies and the model in terms of e.g. geometry, duration of the ram-pressure event and ICM density.

The SFR in the tail of the model of \citet{tonnesen12} starts from 0, to reach $\sim$ 0.05 M$_{\odot}$ yr$^{-1}$ after about 250 yr of
ram-pressure wind. The stellar mass accumulated during this period reaches about 40 10$^5$ M$_{\odot}$. These numbers are larger than found in our results by at least one order of magnitude (see Table \ref{tabtailq}), suggesting either that our galaxies have felt the effects of the wind for a shorter time, or that star formation is less
efficient in our galaxies than in the model. 
 
The average surface densities of (young) stars in the tails which we derive from the UV, $log(\Sigma_*)$, are about 10$^2$ to 10$^3$ M$_{\odot}$
kpc$^{-2}$. Fig. 13 of \citet{tonnesen12} shows that their model produces a large area with stellar surface densities above 10$^3$, reaching peak values of 10$^4$ M$_{\odot}$ kpc$^{-2}$ or more, which appears rather higher than our estimates.

In summary, the observations appear to indicate that the present models tend to overestimate the amount of star formation in the tails, at least under the conditions prevalent in the Virgo cluster where large \HI\ tails are present. The comparison between the data and the models may not tell the full story since, beyond the differences discussed above, part of the stripped gas may not be observed at present, as it was lost by mixing and evaporation \citep{vollmer07}, two phenomena not included in the models.
While the comparison is still rather qualitative at present, the numbers given in Table \ref{tabtailq} could be used as constraints for future generations of models, at least for conditions such as found in the Virgo cluster.

\section{Conclusions}
 \label{sec:conclu}

We analyzed the star formation associated with \HI\ tails due to ram-pressure stripping in 8 galaxies in the Virgo cluster. 
We inspected FUV GALEX images from the GUViCS survey, as well as deep optical images from the NGVS survey. We performed a pixel-based analysis of the relation between the FUV surface brightness and the \HI\ surface density (i.e., one of the versions of the so-called Schmidt Law). As a global test, we also computed the mass fractions of newly formed stars and \HI\ gas found outside the optical body of the galaxies, and global properties of the tails. These quantities were compared to models from the literature which include star formation in ram-pressure stripped gas.
Our findings are:
\begin{itemize}

\item In two cases, \object{NGC 4330} and \object{NGC 4522}, FUV emission is seen in the \HI\ tails, and even blue clusters/filaments are seen in the optical images of the tails, though only in the part closest to the main body of the galaxy.

\item The same Schmidt Law relation between \HI\ surface density and FUV surface brightness (proxies for respectively the gas and star formation rate density) was found in the outer parts of our galaxies (between 1 and 1.5 $R_{25}$ in radius) as by \citet{bigiel10} for dwarfs and the outer parts of spirals.

\item However, we do not observe this relation in the stripped \HI\ tails. At a given \HI\ surface density, the FUV emission in this external gas is systematically weaker than in the inner and outer parts of the galaxies, suggesting a lower SFR in the tails. In fact, our FUV measurements can be considered as upper limits and are in most cases close to the detection limit.

\item The mass fraction of newly formed stars outside a galaxy (estimated in the FUV) compared to stars inside the galaxy is systematically much smaller than the \HI\ mass fraction in the tails.

\item Our results suggest that the star formation efficiency in gas stripped from galaxy by ram-pressure is lower by a factor of at least 10 than within galaxies. This low efficiency may be related to the fact that the stripped gas has disturbed kinematics (diverging flows, increased velocity dispersion), and to the absence of confinement by the gravitational potential of the galactic disk \citep{vollmer12sfe}. While within galactic disks, the stellar density may contribute to star formation through its potential \citep[e.g.][]{abramova08,leroy08}, the stripped gas is less likely to become dense enough to form stars, and more likely to disperse in the relatively low pressure of the ICM gas in the Virgo cluster.

\item In other clusters there are, however, indications of star formation likely associated with ram-pressure stripping events \citep{cortese07,smith10,sun10,yoshida12}. This difference could be related to different physical conditions in the samples. For instance, the IGM of Virgo is about 10 times less dense and hot than in rich clusters such as Coma. Higher densities and pressure in a cluster could help the gas confinement and star formation in rich clusters. Our selection of galaxies with large \HI\ tails could also have prevented us from finding young stars if star formation in stripped tails were abrupt and violent: either \HI\ is observed before stars are formed in it (our selection), or 
the stars have already formed but all the gas has been dispersed or ionized.

\end{itemize}

The physics of star formation is extremely complex (turbulence, cooling, pressure, feedback...), and even more so under the exceptional conditions encountered in gas stripped away from galaxies by ram-pressure \citep[where additional processes could take place, such as thermal evaporation, see][]{cowie77}. Attempts to model its physics have nevertheless been undertaken by several teams \citep[e.g.][]{kapferer09,tonnesen12}. Comparisons between such models and observational data often remain relatively qualitative. However, our study of Virgo cluster galaxies for which a recent ram-pressure stripping event is well established, and where evidence for tidal interaction is absent, provides sufficiently robust constraints on the amount of star formation in the tails (and its relation to the \HI\ surface density) to be taken into account by a future generation of models.

\begin{acknowledgements}

We thank B. Vollmer for discussions that formed the basis of this project and for his critical comments, and the referee for detailed comments and suggestions. This work is supported in part by the French Agence Nationale de la Recherche (ANR) Grant Programme Blanc VIRAGE 
(ANR10-BLANC-0506-01). 
The research leading to these results has received funding from the European Community's Seventh Framework Programme (FP7/2007-2013) under grant agreement No. 229517.
This work is supported in part by the Canadian Advanced Network for Astronomical Research (CANFAR) which has been made possible by funding from CANARIE under the Network-Enabled Platforms program. This research also used the facilities of the Canadian Astronomy Data Centre operated by the National Research Council of Canada with the support of the Canadian Space Agency.
This research has made use of the GOLD Mine Database, and the NASA/IPAC Extragalactic Database (NED) which is operated by the Jet Propulsion Laboratory, California Institute of Technology, under contract with the National Aeronautics and Space Administration. 
We wish to thank the GALEX Time Allocation Committee for the generous allocation of time devoted to GUViCS, and the Canada-France-Hawaii Telescope direction and staff for the support that enabled the NGVS.
GALEX is a NASA Small Explorer, launched in 2003 April. We gratefully acknowledge NASA's support for the construction, operation and science analysis for the GALEX mission, developed in cooperation with the Centre National d'Etudes Spatiales of France and the Korean Ministry of Science and Technology.

\end{acknowledgements}

\appendix

\section{Discussion of individual galaxies}

\label{sec:ind}

We comment on each galaxy on the basis of the GALEX FUV images (Fig. \ref{FigUVdata}), the optical NGVS images (Fig. \ref{FigNGVSdata}) and their individual \HI-FUV relationship as shown in on-line figures \ref{FigLocalN4294} to \ref{FigLocalN4522}. In these figures, individual pixels above the solid (dotted) lines are detected at the 3 (1) $\sigma$ level, respectively. Although most points in the \HI\ tails are barely detected in the FUV, we keep all pixels in our analysis, in particular because we compute the average FUV surface brightness in \HI\ surface density bins from many pixels, thus reducing the noise level. In our figures, the point near the lower-right corner indicates the FUV intensity obtained by stacking in the same manner all pixels without detected \HI, which provides an indication of the FUV emission detection limit for the binned analysis.

\onlfig{1}{
\begin{figure*}
\includegraphics[width=0.95\textwidth]{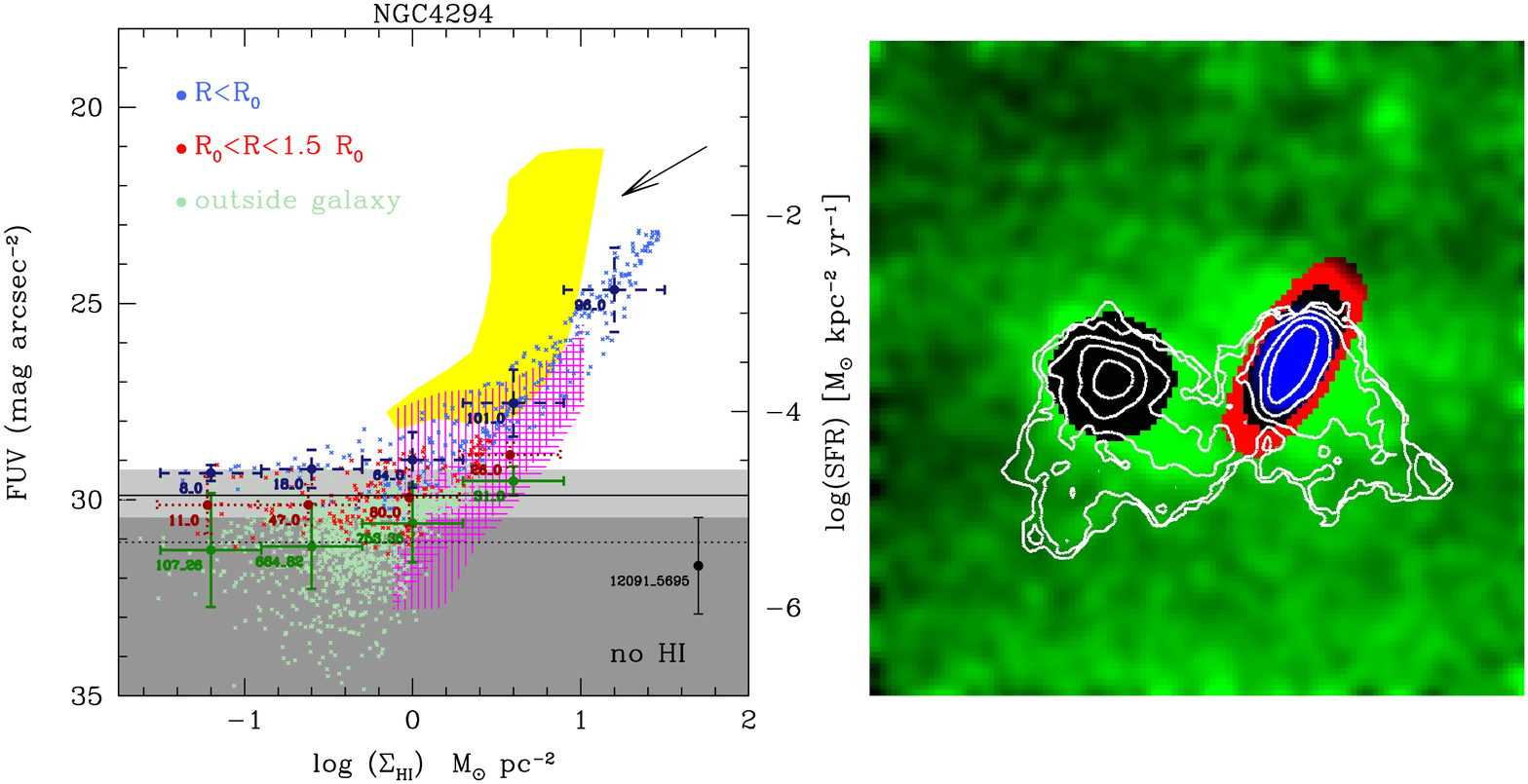}
\caption{On the left, the relationship between FUV surface brightness and \HI\ surface density is shown for \object{NGC 4294}.
The colors indicate which of the three defined regions the pixels belong to: inner galaxy in blue, outer galaxy in red and external gas in green (as shown in the right-hand panel). The 1 (3) $\sigma$ FUV sky surface brightness level is indicated by the horizontal dotted (solid) line.
The points with the error-bars indicate, from top to bottom, the mean and its dispersion within the \HI\ surface density bins for each of the three regions (inner galaxy, extended galaxy and external gas).
We have not applied any inclination corrections to these data. However, the arrow indicates how the pixel values should be corrected to account for inclination (based on the major/minor axis ratio of the galaxy). The correction in the FUV is very uncertain however, because it assumes the galaxy is optically thin. 
The two numbers on the left of each error bar, separated by an underscore symbol, indicate (left) the number of pixels used for this point, and (right) the number of pixels with a flux below the mean sky level in the bin. As using the latter pixels would give an infinite surface brightness they are not included in the computation of mean surface brightness. A large value for this second number is a good indication that we are very close to the sky value. The error bar near the lower-right corner (labeled ``no \HI'') shows the result of the same analysis when performed on the FUV emission outside the detected \HI\ only, and provides an indication of our FUV detection limit (the dark- and light- grey shaded areas show, respectively, the corresponding 1 and 2 $\sigma$ limits). 
The yellow-shaded area indicates the location of nearby field galaxies from the THINGS sample \citep{bigiel08}. The vertically and horizontally shaded areas correspond respectively to dwarf galaxies and the outer parts of spirals, i.e., the points with error bars in Fig. 8 of \citet{bigiel10}. A similar diagram for our complete sample is shown in our Fig. \ref{FigSLcomb}. 
}     
\label{FigLocalN4294}
   \end{figure*}
}

\onlfig{2}{
\begin{figure*}
\includegraphics[width=0.95\textwidth]{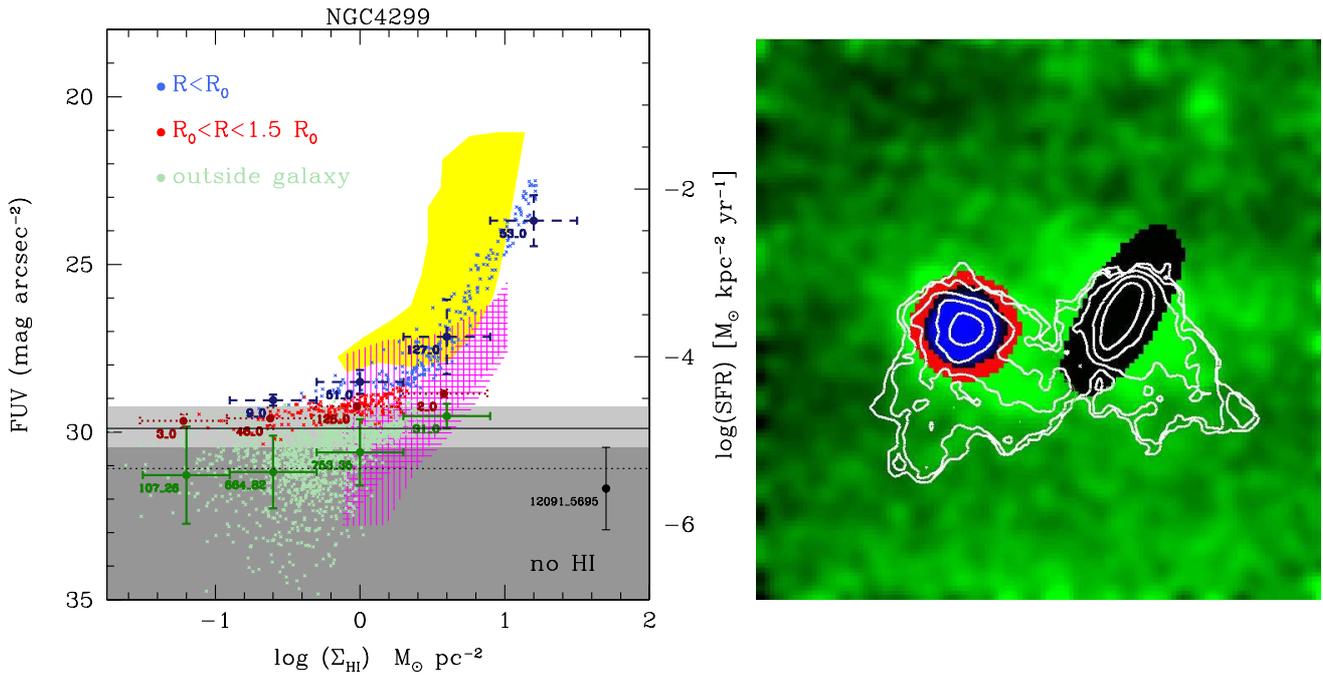}
\caption{FUV-\HI\ relationship in \object{NGC 4299}. Same as Fig. \ref{FigLocalN4294}}         
\label{FigLocalN4299}
   \end{figure*}
}

\onlfig{3}{
\begin{figure*}
\includegraphics[width=0.95\textwidth]{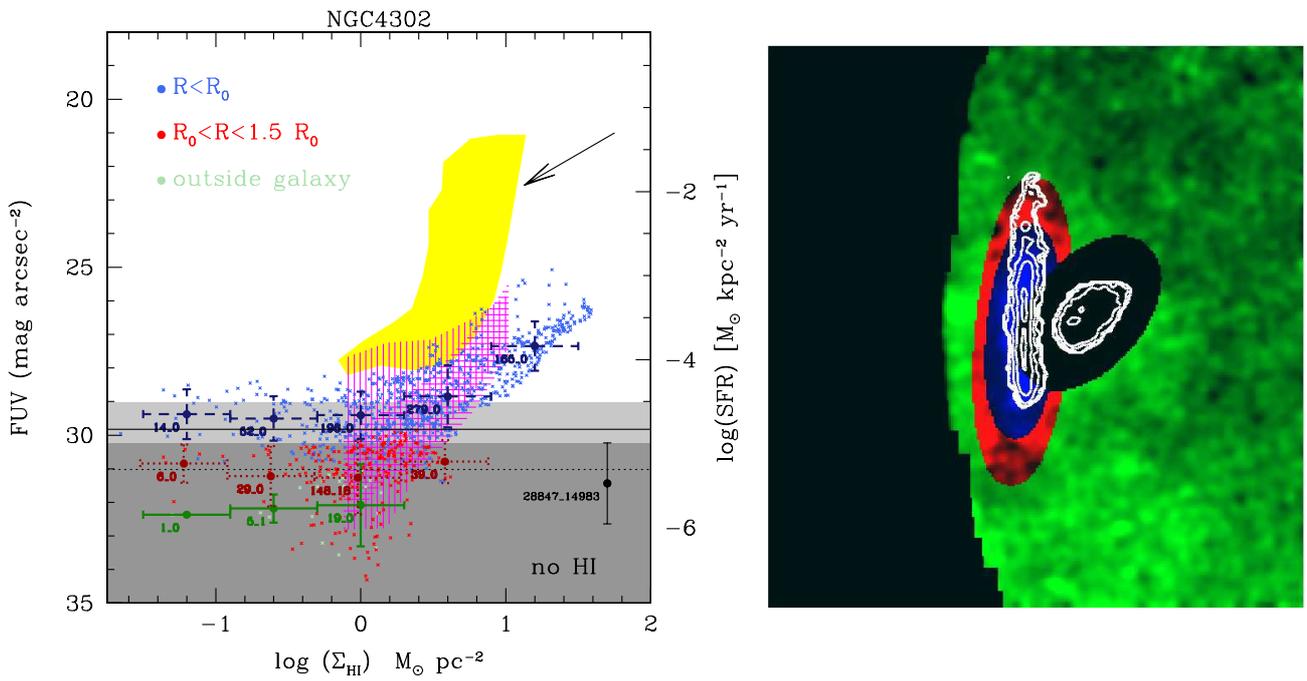}
\caption{FUV-\HI\ relationship in \object{NGC 4302}. Same as Fig. \ref{FigLocalN4294}}         
\label{FigLocalN4302}
   \end{figure*}
}

\onlfig{4}{
\begin{figure*}
\includegraphics[width=0.95\textwidth]{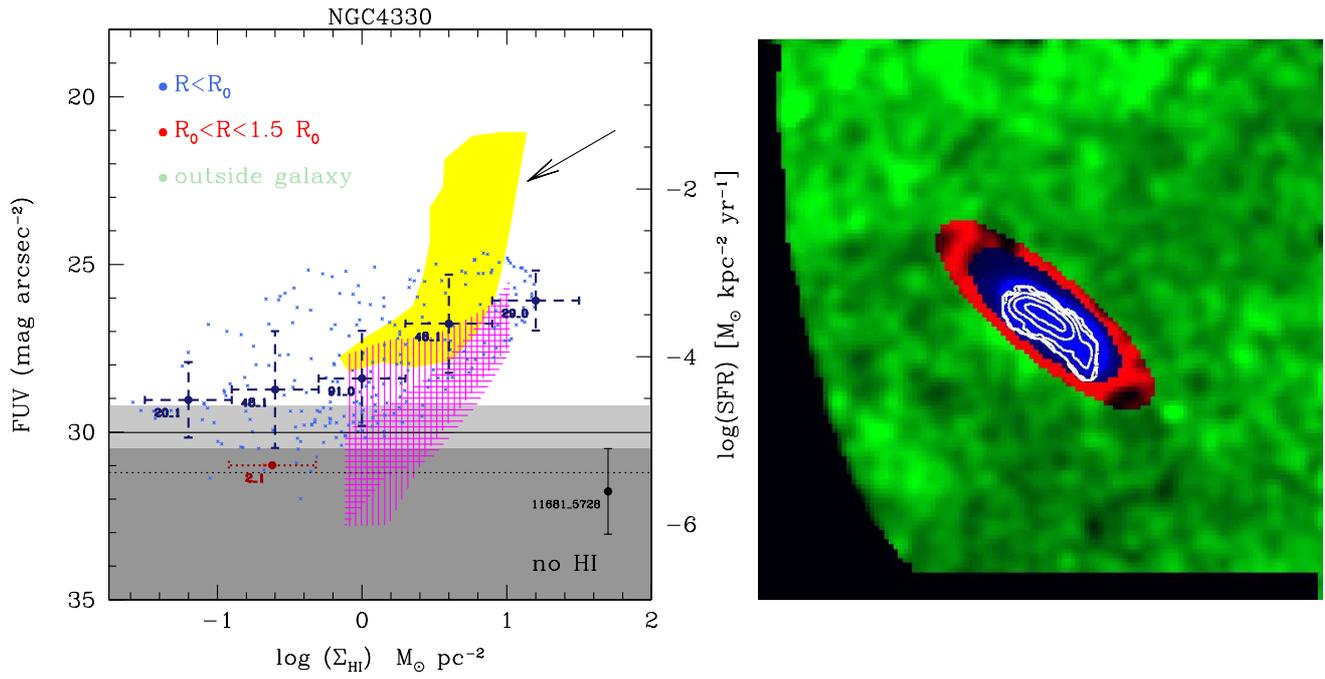}
\caption{FUV-\HI\ relationship in \object{NGC 4330}. Same as Fig. \ref{FigLocalN4294}}         
\label{FigLocalN4330}
   \end{figure*}
}

\onlfig{5}{
\begin{figure*}
\includegraphics[width=0.95\textwidth]{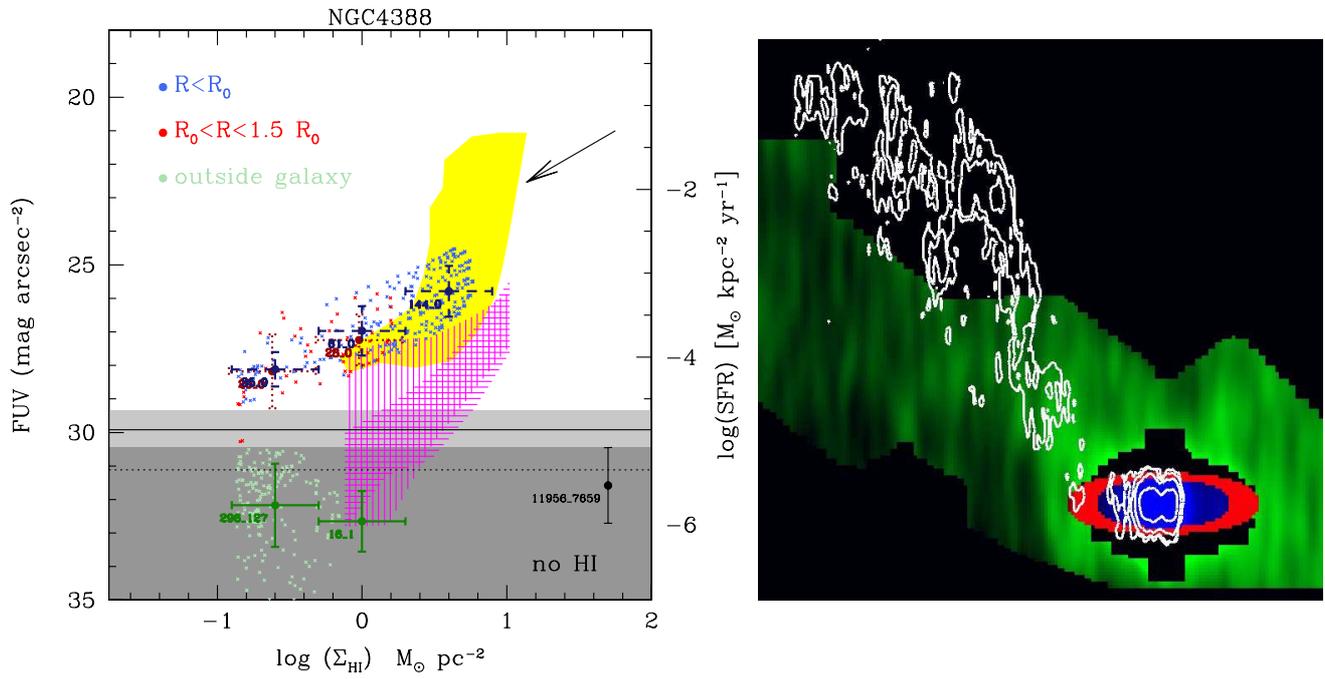}
\caption{FUV-\HI\ relationship in \object{NGC 4388}. Same as Fig. \ref{FigLocalN4294}}         
\label{FigLocalN4388}
   \end{figure*}
}

\onlfig{6}{
\begin{figure*}
\includegraphics[width=0.95\textwidth]{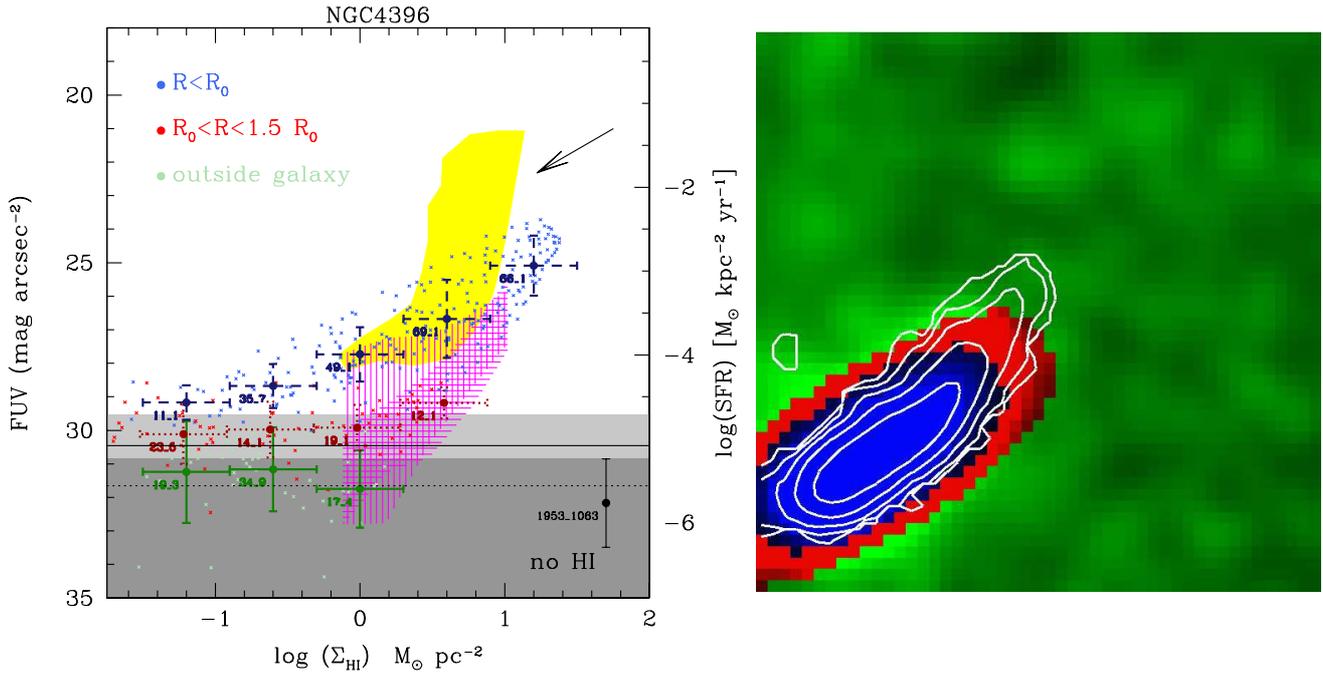}
\caption{FUV-\HI\ relationship in \object{NGC 4396}. Same as Fig. \ref{FigLocalN4294}}         
\label{FigLocalN4396}
   \end{figure*}
}

\onlfig{7}{
\begin{figure*}
\includegraphics[width=0.95\textwidth]{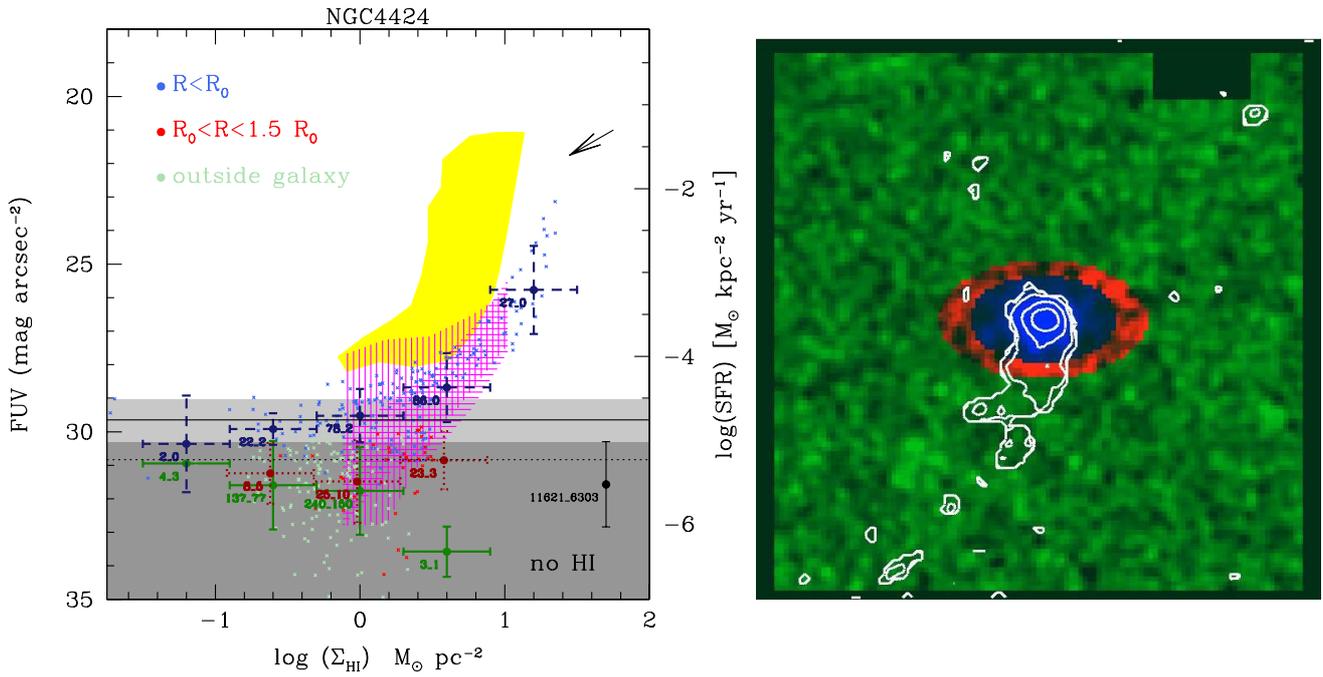}
\caption{FUV-\HI\ relationship in \object{NGC 4294}. Same as Fig. \ref{FigLocalN4294}}         
\label{FigLocalN4424}
   \end{figure*}
}

\onlfig{8}{
\begin{figure*}
\includegraphics[width=0.95\textwidth]{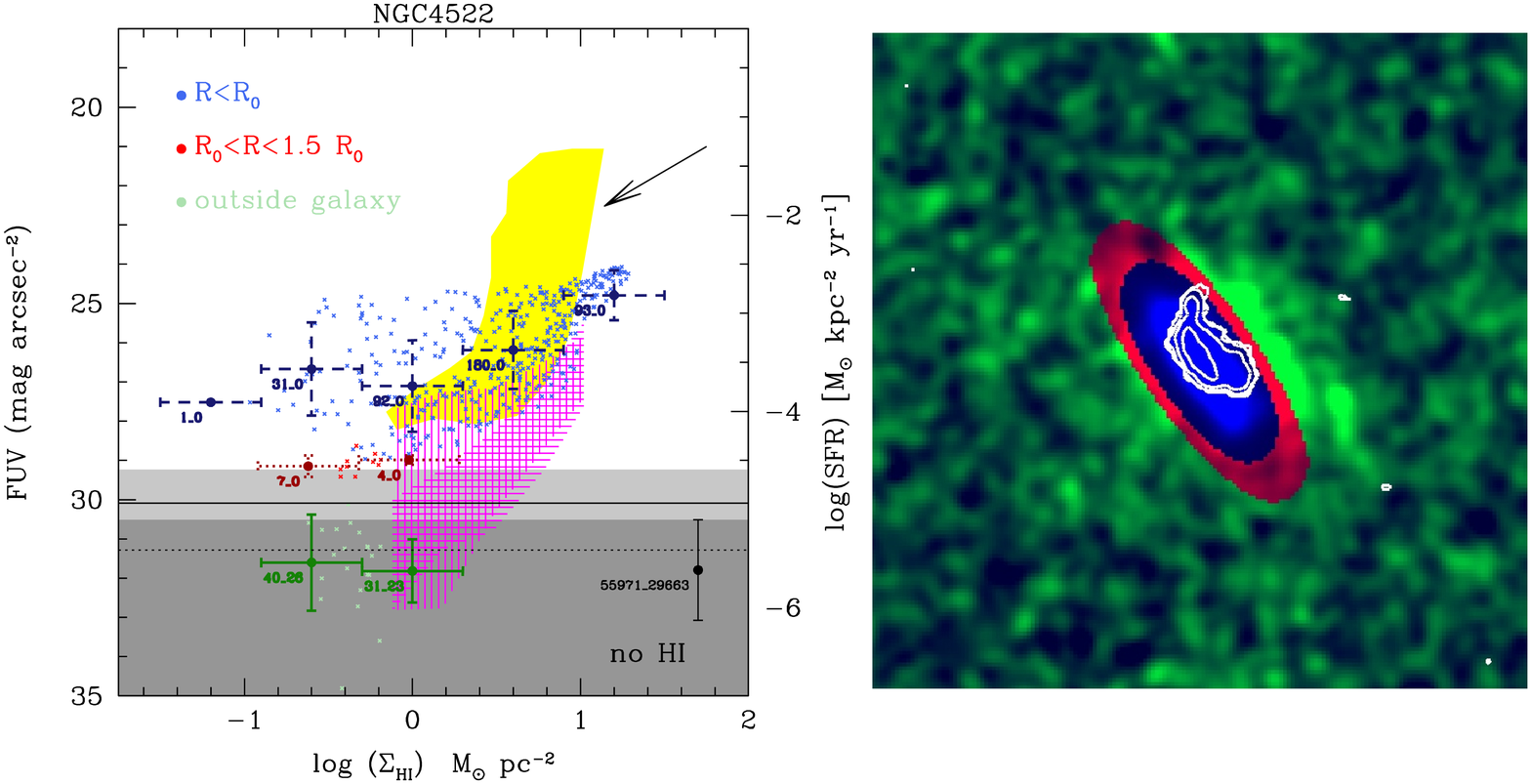}
\caption{FUV-\HI\ relationship in \object{NGC 4522}. Same as Fig. \ref{FigLocalN4294}}         
\label{FigLocalN4522}
   \end{figure*}
}

\subsection{NGC 4294 and NGC 4299}

\citet{chung09} found no optical counterpart in SDSS images of the \HI\ tail down to $r=$26 mag arcsec$^{-2}$, which we confirm at the much lower NGVS detection limit \citep[about 29 mag arcsec$^{-2}$ in the $g$-band at the 2 $\sigma$ level,][]{laurasub}.
They also mention that the H$\alpha$ emission is enhanced in the south of \object{NGC 4299}. This enhancement is also seen in our FUV images. It is however found within the optical radius of the galaxy only, and no clear UV clusters are seen outside the galaxy in the extended \HI\ tail shared by \object{NGC 4294} and \object{NGC 4299}. In our on-line Figs. \ref{FigLocalN4294} and \ref{FigLocalN4299}, we have treated all external gas together, rather than making a separate analysis for each galaxy -- in any case, the gas tail is almost common to both galaxies.
For both galaxies, the inner galaxy \HI-SFR relationship is in agreement with the data on spirals and dwarfs from \citet{bigiel08} at surface densities around a few M$_{\odot}$ yr$^{-1}$; we remind the reader that a comparison at larger densities is difficult because of the increasing dust attenuation and molecular fraction). The pixels in the outer regions of the two galaxies fall in the \citet{bigiel10} area for dwarfs and outer regions of spirals, as expected. The external FUV emission level (in the tail) is well below the FUV emission within the galaxy (inner or outer) at any given \HI\ surface density. The tail's FUV emission is moreover very close to the detection limit (a few sigma).

\subsection{NGC 4302}

Note that \object{NGC 4298} and \object{NGC 4302} form a pair (like \object{NGC 4294} and \object{NGC 4299}). While the interaction may affect the level and distribution of gas and star formation within \object{NGC 4298}, it is not believed to be undergoing ram-pressure stripping, and is therefore not included in the study.
\object{NGC 4302} has no UV emission associated with the \HI\ tail to the north. The rest of the \HI\ disk is truncated within the optical radius of the galaxy. A faint level of diffuse emission is seen in NGVS images at the position of the \HI\ tail, but it does not show clear signs of young clusters or filaments, and it has similar levels of emission and colors as were found on the opposite side of the galaxy, where no gas is present. No FUV emission is seen in the tail, suggesting a low star formation efficiency in the external gas. In our pixel-by-pixel analysis, however, too few pixels are located outside of the optical body of the galaxy to derive any useful constraints.

\subsection{NGC 4330}

\citet{chung07} found a UV tail along the edge of the \HI\ to the southwest of the galaxy, displaced from the optical. Our FUV image also shows this extension (Fig. \ref{FigUVdata}). \citet{chung07} did not see any optical counterpart to the \HI\ on SDSS images. However, we clearly see blue clusters in our deeper NGVS images (Fig. \ref{FigNGVSdata}) corresponding to the southwest FUV emission, but slightly displaced from it.
The recent detailed study by \citet{abramson11} also found H$\alpha$ and FUV emission, suggesting a low level of star formation in the tail, and confirmed the offset between the newly formed stars and the gas, which can occur if the ISM is continuously accelerated by ram-pressure, while stars decouple from it \citep{vollmer12ngc4330}.
They also discussed nine extra-planar star forming regions found in the UV southeast of the galaxy. They estimated their ages to be between 100 and 350 Myr, from their FUV-NUV colors. These regions are not associated with gas peaks. 

The gas disk of the galaxy is severely truncated, and the optical disk is much larger. For this reason, the young stars' emission discussed above is found entirely ``within'' the galaxy, according to our definition of regions (section \ref{sec:tagging}), especially at the \HI\ resolution used for studying the \HI-FUV relationship (our pixel analysis actually has no points ``outside'' the galaxy, see Fig. \ref{FigLocalN4330}). 

Some of the extra-planar star-forming regions are also detected as blue clusters in our NGVS images (Fig. \ref{FigNGVSdata}). They are clearly found close to the galaxy only and not over the full extent of the external \HI, as are the UV clusters, suggesting again a relative low star formation efficiency in the \HI. A detailed analysis of the stellar population based on the NGVS data is beyond the scope of this paper.

\subsection{NGC 4388}
\label{sec:ind4388}

This galaxy is known for its very long \HI\ tail, probably made of stripped gas \citep{vollmer03,oosterloo05,chung09}. We see no obvious optical or FUV counterparts in the tail, and the GALEX data are consistent with the absence of UV emission over the tail. Note that the optical/UV analysis cannot be performed over the full area of the tail because of the proximity of bright M86, which is masked in our pixel-based analysis.
Because of the active nucleus of the galaxy (very bright in the UV) and of the very elongated shape of the WSRT \HI\ beam (and thus of the Gaussian kernel used to attain this resolution with the UV data), part of the flux from ``inside'' the galaxy is found ``outside'' it, in our definition of the various regions. To avoid counting this flux as ``external'', we had to mask it as much as possible. Although this masking (visible in the right part of Fig. \ref{FigLocalN4388}) introduces a small error in the \HI\ and UV content of the galaxy, it avoids over-estimating the surface brightness and density values outside the galaxy.

Finally, we wish to remind that this work is performed at the \HI\ resolution, so we cannot distinguish fine details within the disk (nor is this our purpose, as we try to estimate the amount of star formation in the tail, outside the optical body). However, some optical emission coinciding with H$\alpha$ regions very close to the disk of the galaxy was found in NGVS images by Ferri\`ere et al. (in preparation). While \citet{yoshida04} attributed the H$\alpha$ emission along the \HI\ tail to ionization of the stripped gas by an active nucleus, Ferri\`ere et al. discuss the possibility that some of these regions, very close to the disk, may be associated with star formation.

\subsection{NGC 4396}

\citet{chung07} did not note any UV emission or optical emission in the SDSS image which may be linked to the extended \HI. Neither do we see anything in our GALEX UV images, nor blue clusters in the NGVS optical data.
In the pixel-by-pixel analysis, very few pixels are in fact outside the galaxy (Fig. \ref{FigLocalN4396}) and they show a very low level of FUV emission.

\subsection{NGC 4424}

Inside the galaxy, the FUV emission is relatively faint: it is consistent with the outer regions of spirals and with dwarfs, rather than the inner parts of normal spirals. The FUV image shows no wide-spread star formation in the gas outside the galaxy (Fig. \ref{FigLocalN4424}) and no obvious filaments/clumps are seen in the NGVS images.

\subsection{NGC 4522}

\object{NGC 4522} is undergoing a strong ram-pressure event despite its large distance to the center of the cluster \citep{kenney04,vollmer06,chung09}.
\citet{kenney99} and \citet{kenney04} showed the presence of \HII{} regions in the extra-planar gas. We indeed see some FUV emission when going out towards the extra-planar \HI\ gas peaks, but none at the radius of the outermost isophote where \HI\ is detected. 
The NGVS optical images also reveal blue filamentary structures and clusters in the same direction, which are barely visible in the $B$-band image of \citet{kenney99}.
Once smoothed to the \HI\ resolution, almost no pixels with detected \HI\ are found beyond $R_{25}$. Fig. \ref{FigLocalN4522} shows that the only contribution to the extended galaxy region (at $R_{25}<R<1.5R_{25}$) is by this extra-planar gas. The FUV emission level in the extended region is clearly lower than in the inner galaxy, which itself seems, on the contrary, higher than the typical value in the THINGS sample \citep{bigiel08} at the same gas surface density. 
\citet{crowl06} found that the outer disk of the galaxy is consistent with a 2\% (in stellar mass) very short duration starburst which occurred at the epoch of the truncation (100 Myr ago). The observed UV emission and blue filaments in the immediate proximity of the galaxy could correspond to stars formed during this brief episode.

\end{document}